\documentclass[11pt]{article}
\pdfoutput=1
\usepackage{jcapmod}

\usepackage{booktabs}
\usepackage[english]{babel}
\usepackage[per-mode=reciprocal, mode=math, group-digits=integer]{siunitx}[=v2]
\usepackage[babel=true]{microtype}

\numberwithin{equation}{section}
\allowdisplaybreaks[1]

\def\d{\mathrm{d}}
\def\ee{\mathrm{e}}
\def\ii{\mathrm{i}}

\def\k{\vec{k}}
\def\p{\vec{p}}
\def\q{\vec{q}}
\def\x{\vec{x}}
\def\O{\mathcal{O}}
\def\T{\mathcal{T}}

\def\fnl{f_\mathrm{NL}}
\def\fnlloc{f_\mathrm{NL}^\mathrm{loc}}
\def\fnleq{f_\mathrm{NL}^\mathrm{eq}}
\def\fsky{{f_\mathrm{sky}}}
\def\kmax{k_\mathrm{max}}
\def\kmin{k_\mathrm{min}}
\def\knl{k_\mathrm{NL}}
\def\zmax{z_\mathrm{max}}
\def\zmin{z_\mathrm{min}}

\DeclareSIUnit{\parsec}{pc}
\DeclareSIUnit{\Mpc}{\mega\parsec}
\DeclareSIUnit{\Gpc}{\giga\parsec}
\DeclareSIUnit{\h}{\mathit{h}}
\DeclareSIUnit{\hPerMpc}{\h\per\Mpc}
\DeclareSIUnit{\MpcPerh}{\per\h\Mpc}

\DeclareRobustCommand{\SkipTocEntry}[4]{}

\begin{document}
	
\pagenumbering{roman}
\begin{titlepage}
	\baselineskip=15.5pt \thispagestyle{empty}
	
	\bigskip\
	
	\vspace{1cm}
	\begin{center}
		{\fontsize{20.74}{24}\selectfont \sffamily \bfseries Light Fields during Inflation\\[8pt]from BOSS and Future Galaxy Surveys}
	\end{center}
	
	\vspace{0.2cm}
	\begin{center}
		{\fontsize{12}{30}\selectfont Daniel Green,$^{\bigstar}$ Yi Guo,$^{\bigstar}$ Jiashu Han\hskip1pt$^{\bigstar}$ and Benjamin Wallisch\hskip1pt$^{\spadesuit,\bigstar,\blacklozenge,\clubsuit}$}
	\end{center}
	
	\begin{center}
		\vskip8pt
		\textsl{$^\bigstar$ Department of Physics, University of California San Diego, La Jolla, CA 92093, USA}
		
		\vskip8pt
		\textsl{$^\spadesuit$ Oskar Klein Centre, Department of Physics, Stockholm University, 10691~Stockholm, Sweden}
		
		\vskip8pt
		\textsl{$^\blacklozenge$ School of Natural Sciences, Institute for Advanced Study, Princeton, NJ 08540, USA}
		
		\vskip8pt
		\textsl{$^\clubsuit$ Center for Cosmology and Astroparticle Physics, Weinberg Institute for Theoretical Physics,\\Department of Physics, The University of Texas at Austin, Austin, TX~78751, USA}
	\end{center}

	\vspace{1.2cm}
	\hrule \vspace{0.3cm}
	\noindent {\sffamily \bfseries Abstract}\\[0.1cm]
Primordial non-Gaussianity generated by additional fields present during inflation offers a compelling observational target for galaxy surveys. These fields are of significant theoretical interest since they offer a window into particle physics in the inflaton sector. They also violate the single-field consistency conditions and induce a scale-dependent bias in the galaxy power spectrum. In this paper, we explore this particular signal for light scalar fields and study the prospects for measuring it with galaxy surveys. We find that the sensitivities of current and future surveys are remarkably stable for different configurations, including between spectroscopic and photometric redshift measurements. This is even the case at non-zero masses where the signal is not obviously localized on large scales. For realistic galaxy number densities, we demonstrate that the redshift range and galaxy bias of the sample have the largest impact on the sensitivity in the power spectrum. These results additionally motivated us to explore the potentially enhanced sensitivity of Vera Rubin Observatory's~LSST through multi-tracer analyses. Finally, we apply this understanding to current data from the last data release of the Baryon Oscillation Spectroscopic Survey~(BOSS~DR12) and place new constraints on light fields coupled to the inflaton.

	\vskip10pt
	\hrule
	\vskip10pt
\end{titlepage}

\thispagestyle{empty}
\setcounter{page}{2}
\tableofcontents

\clearpage
\pagenumbering{arabic}
\setcounter{page}{1}
\section{Introduction}
\label{sec:introduction}

The statistics of the primordial density perturbations offer a window into the dynamics of the very early universe~\cite{Green:2022hhj, Chang:2022lrw, Green:2022bre}, in particular the inflationary epoch~\cite{Flauger:2022hie, Achucarro:2022qrl}. While current maps of the universe are consistent with purely Gaussian fluctuations~\cite{Planck:2019kim}, non-Gaussian correlation functions encode the particles and interactions relevant to the origin of structure~\cite{Meerburg:2019qqi, Biagetti:2019bnp, Achucarro:2022qrl}. Inflation predicts a lower bound on the amount of primordial non-Gaussianity~(PNG) due to gravity alone that is approximately two orders of magnitude below current constraints~\cite{Cabass:2016cgp}. On the other hand, signals that are large enough to be detected in the next generation of surveys arise in many models, including examples with only Planck-suppressed interactions~\cite{Assassi:2013gxa}.\medskip

Due to the limited number of modes remaining to be measured to the cosmic variance limit in the cosmic microwave background~(CMB), surveys of the large-scale structure~(LSS) of the universe present the best opportunity to improve our understanding of the primordial statistics~\cite{Achucarro:2022qrl, DESI:2022lza, Schlegel:2022vrv, Annis:2022xgg, Chang:2022lrw}. With the benefit of three-dimensional information, even current surveys have the raw statistical power to compete with the CMB~\cite{Beutler:2019ojk}. Unfortunately, for generic non-Gaussian correlation functions, late-time nonlinearities make many modes inaccessible and present a serious obstacle to our progress~\cite{Alvarez:2014vva, Baldauf:2016sjb, Cabass:2022avo, Cabass:2022epm}. In some circumstances, however, the manifestation of the primordial signal is robust to nonlinear physics and can be measured reliably~\cite{Green:2022hhj}.\medskip

The fluctuations of extra fields beyond the inflaton present such an opportunity. In the absence of these fields, correlation functions of matter and galaxies are subject to the constraints of the single-field consistency conditions~\cite{Creminelli:2004yq, Creminelli:2013mca, Creminelli:2013poa}. In practice, these conditions are similar to the equivalence principle~\cite{Creminelli:2013nua} and ensure that correlations are determined by the curvature of spacetime (i.e.~derivatives of the metric fluctuations). Light fields break these conditions and can lead to enhanced long-distance correlations~\cite{Lyth:2002my, Zaldarriaga:2003my, Sasaki:2006kq, Chen:2009we, Chen:2009zp, Baumann:2011nk}. These apparent violations of the equivalence principle cannot arise from nonlinear gravity, and can therefore be distinguished from astrophysical and gravitational sources of non-Gaussianity.\medskip

\begin{figure}
	\centering
	\includegraphics{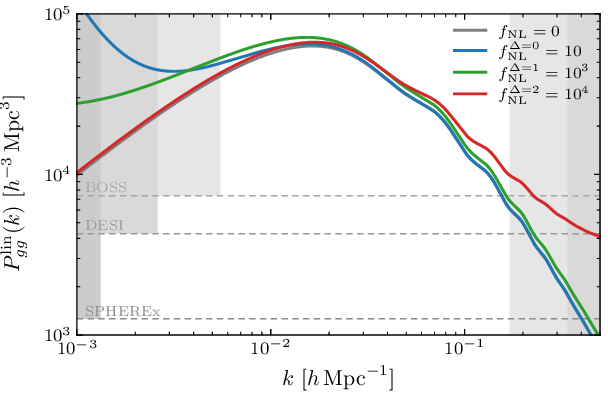}\vspace{-7pt}
	\caption{Illustration of the effect of the scale-dependent bias~$b_\mathrm{NG}^\Delta$ from a non-zero non-Gaussian amplitude~$\fnl^\Delta$ on the linear galaxy power spectrum. We see that the enhancement of power shifts from large to small scales as we increase the exponent~$\Delta$~(see~\textsection\ref{sec:bias} for its definition and details). The horizontal dashed lines indicate the effective shot noise level for~BOSS, DESI and~SPHEREx after scaling them to the displayed redshift~$z=0$ and linear bias~$b_1 = 1.6$. The gray-shaded regions on large scales indicate the wavenumbers below the minimum wavenumber of these surveys, $k < \kmin$, computed based on their entire spherical volume. The gray-shaded regions for large wavenumbers indicate the regimes where the scales exceed the nonlinear scale, $k > \knl$, for the maximum redshift of~BOSS and~DESI, respectively.}
\label{fig:bias}
\end{figure}

From the perspective of both data analysis and survey design, the unique advantage of these non-Gaussian signals is that they can be observed in the two-point statistics of galaxies. Violations of the single-field consistency conditions introduce couplings between long- and short-wavelength modes. The short-wavelength modes control the number density of galaxies, which are then non-trivially correlated on large scales in the presence of this type of non-Gaussianity. The resulting~(non-local) change to the shape of the galaxy power spectrum, shown in Figure~\ref{fig:bias},%
is known as scale-dependent bias~\cite{Dalal:2007cu} and has been searched for in many existing dataset~(see e.g.~\cite{Slosar:2008hx, Ross:2012sx, Leistedt:2014zqa, Castorina:2019wmr, Mueller:2021jvn, DAmico:2022gki, Cabass:2022ymb, Rezaie:2023lvi, DESI:2023duv, Cagliari:2023mkq}).\medskip

The case of local primordial non-Gaussianity~\cite{Komatsu:2001rj}, for which the Gaussian and non-Gaussian Newtonian potentials~$\varphi$ and~$\Phi$ are related by
\begin{equation}
	\Phi(\x) = \varphi(\x) - \fnlloc \varphi^2(\x) + \ldots\, ,	\label{eq:local-png}
\end{equation}
has been very well studied in the literature, both theoretically and observationally. Local~PNG arises in the presence of multiple massless fields, such that~$\varphi$ is an isocurvature mode during inflation, but is converted to an adiabatic mode at later times~\cite{Lyth:2002my, Zaldarriaga:2003my, Sasaki:2006kq}. Future measurements are expected to reach particularly interesting thresholds for the physics of multi-field inflation~\mbox{\cite{Alvarez:2014vva, SPHEREx:2014bgr, dePutter:2016trg, Achucarro:2022qrl}.}

The contribution to primordial non-Gaussianity from massive particles is much less studied observationally. It is however a classic signature of quasi-single-field inflation~\cite{Chen:2009we, Chen:2009zp, Baumann:2011nk}, which is also known as cosmological collider physics~\cite{Arkani-Hamed:2015bza}. Importantly, these models are compelling targets for current and future surveys. In spite of this, little has been known about the optimal survey strategy to search for these relics from inflation up to now.\medskip

In this paper, we will examine how future surveys can best constrain these non-Gaussian signals using galaxy power spectra. For local~PNG~(massless fields), the signal is dominated by the largest scales and most observational strategies are designed accordingly~\cite{dePutter:2014lna, Mueller:2017pop}. On the other hand, the signals of massive fields may arise at large or small scales~(see Fig.~\ref{fig:bias}), which means that the characteristics which best constrain these models are less clear a~priori~\cite{Sefusatti:2012ye, Norena:2012yi, Gleyzes:2016tdh}. We therefore study this signal in detail, and investigate how redshift coverage, biasing, multi-tracer techniques and target selection impact forecasts over the full range of scaling laws induced by these particles.

We find that target selection, specifically finding highly biased objects, is the largest factor in driving current and future measurements of this effect when assuming a fixed universality relation. While large volumes help to increase the number of modes, especially in spectroscopic surveys, we observe that the benefits of going to higher redshifts are more driven by the large biases of high-redshift objects and only secondarily by the increased volume. In photometric surveys, the larger number density of galaxies also enables a significant increase in sensitivity through sample-variance cancellation. This requires splitting the sample according to the bias, which we explore in the context of Vera Rubin Observatory's Legacy Survey of Space and Time~(LSST).

Using these insights, we explore current constraints on the scale-dependent bias of galaxies using the BOSS~DR12 dataset~\cite{BOSS:2016wmc} for the full range of scaling behaviors of light inflationary fields. We measure the contributions to the power spectrum for one scaling exponent~(PNG~shape) at a time and compute a correlation matrix to extrapolate between the discrete choices of scaling dimensions~(masses). In terms of the amplitude~$\fnl^\Delta$ of primordial non-Gaussianity~(and, therefore, of the scale-dependent bias), we infer the following discrete results at~$2\sigma$:
\begin{equation}
	\fnl^\Delta = \left\{1_{-39}^{+35},\, 0_{-300}^{+260},\, 400_{-2500}^{+2900},\, 250_{-3900}^{+4000},\, 200_{-5300}^{+5400}\right\}\quad\text{ for }\quad \Delta = \{0.0,\, 0.5,\, 1.0,\, 1.5,\, 2.0\}\, ,	\label{eq:boss-constraints}
\end{equation}
where the parameter~$\Delta$ is related to the mass of the fields, with definitions and details provided below. This means that our analysis is consistent with no scale-dependent bias and zero primordial non-Gaussianity for all light inflationary fields. While our inferred constraints are less sensitive to the PNG~amplitude than those from~Planck, the CMB~constraints for non-zero masses are driven by bispectra in equilateral configurations rather than the scaling behavior in the squeezed limit probed by the galaxy power spectrum. The only previous constraint from LSS~data in this regime was made in~\cite{Agarwal:2013qta}. At the same time, we forecast that galaxy power spectrum measurements in future surveys will exceed the sensitivity of~Planck to these light fields \mbox{for a sizable range of their masses}.\bigskip

This paper is organized as follows: In Section~\ref{sec:background}, we review the relevant theoretical background for our non-Gaussian signal. In Section~\ref{sec:forecasts}, we present forecasts for a wide range of model parameters and experimental configurations. Our goal is to identify the choices that most directly impact the sensitivity to~PNG beyond the local type. In Section~\ref{sec:analysis}, we apply our understanding of the signal to BOSS~DR12~galaxy clustering data and present new constraints on non-Gaussianity from light fields. In Section~\ref{sec:multitracer}, we discuss the role of multi-tracer analyses and astrophysical effects on the forecasts. We conclude in Section~\ref{sec:conclusions}. A set of appendices contains technical details on the modeling of the galaxy power spectrum and the survey specifications~(Appendix~\ref{app:forecasting}), and a discussion on our ability to measure the scaling behavior of the non-Gaussian signature from the galaxy power spectrum~(Appendix~\ref{app:scaling}).

\section{Light Fields and Galaxies}
\label{sec:background}

Our goal in this paper is to explore the signal of additional fields that manifest themselves in the galaxy power spectrum via the scale-dependent bias. This section reviews the necessary background to understand the signal. Readers who are familiar with the scale-dependent bias due to general forms of primordial \mbox{non-Gaussianity~(i.e.\ beyond the local type) may proceed to~Section~\ref{sec:forecasts}.}

\subsection{Galaxy Power Spectrum}
\label{sec:spectrum}

The formation of structure in the universe is driven by the growth of density fluctuations in the dark matter~\cite{Bernardeau:2001qr}. The evolution of the density contrast Fourier mode in redshift space, $\delta_m(\k,z) \equiv \delta \rho_m(\k,z) / \bar \rho_m(z)$, can be solved at linear order to give
\begin{equation}
	\delta_m(\vec{k},z) = \frac{2 k^{2} T(k) D(z)}{3 \Omega_{m} H_{0}^{2}} \Phi(\vec{k}) \equiv k^2\T(k,z) \Phi(\vec{k}) \, ,
\end{equation}
where $k = |\k|$, $T(k)$~is the transfer function defined such that $T(k \to 0) \to 1$, and $D(z)$~is the linear growth factor normalized as $D(z) = 1/(1+z)$ during matter domination. The primordial Newtonian potential~$\Phi(\k)$ encodes the primordial density fluctuations generated during inflation which are concretely of the form
\begin{equation}
	\langle \Phi(\k) \Phi(\k') \rangle = \frac{9}{25} \frac{A_\mathrm{s}}{k^3} k^{n_\mathrm{s}-1} (2\pi)^3\delta_D(\k+\k') \, ,
\end{equation}
where~$\delta_D$ is the Dirac delta function and we used $\Phi = -\frac{3}{5}\zeta$, with curvature fluctuation~$\zeta$, for modes that re-entered the horizon during matter domination. The linear matter power spectrum is therefore given by
\begin{equation}
	P_\mathrm{lin}(k) = k^4\T(k,z)^2 \frac{A_\mathrm{s}}{k^{3-(n_\mathrm{s}-1)}} = \frac{4}{25} \frac{A_\mathrm{s} D(z)^2}{\Omega_m^2 H_0^4}T(k)^2 k^{n_\mathrm{s}} \, .
\end{equation}
The nonlinear matter power spectrum~$P_m(k)$ follows the linear power spectrum on large scales, $P_m(k) \approx P_\mathrm{lin}(k)$ for $k \ll \knl$, for some scale $\knl \approx \SI{0.1}{\hPerMpc}$ at $z=0$. Gravitational evolution corrects this behavior on smaller scales, $k \gtrsim \knl$.\medskip

We will generally work with galaxies in the regime $k < \knl$. The nonlinear effects that are relevant in this regime can be described solely in terms of the bias expansion~\cite{Desjacques:2016bnm, Schmittfull:2018yuk}. In its most general form, the bias expansion is simply the assumption that galaxy formation is a local process:\footnote{Technically, biasing is best understood as a local process in Lagrangian space. The non-locality in Eulerian space is captured by the Zel'dovich approximation~\cite{Schmittfull:2018yuk}.}
\begin{equation}
	\delta_g(\vec{x}) = \sum_{i} b_{\O_i} \O_i(\x) \, ,
\end{equation}
where the operators~$\O_i(\x)$ are any locally measurable quantities. Since we typically assume that the matter controls the formation of galaxies, these are usually local products of $\delta_m(\x)$ or the tidal tensor~\cite{McDonald:2009dh}, $s_{ij} = \left(\frac{\nabla_i \nabla_j}{\nabla^2} - \frac{1}{3}\delta_{ij} \right)\delta_m(\x)$,
\begin{equation}
	\O \in \{ \delta_m, \nabla^2\delta_m, \delta_m^2, s_{ij}^2, \delta_m^3, \delta s_{ij}^2, \mathrm{Tr}[(\Pi^{[1]})^3]|^{(3)}, \mathrm{Tr}[\Pi^{[1]}\Pi^{[2]}]|^{(3)},\cdots \} \, ,
\end{equation}
where the tensors $\Pi^{[n]}_{ij}$ are defined in~\cite{Desjacques:2016bnm}.\footnote{The superscript $(3)$ indicates that the operator includes only terms up to third order in perturbation theory.} In principle, there are an infinite number of operators to consider. At any fixed accuracy, we however understand this as an expansion in powers of the small density contrast, $\delta_m^n \ll 1$, and gradients $R^2 \nabla^2 \ll 1$, for some fixed scale~$R$. To model the galaxy power spectrum at one-loop order, we consider the following set of operators~(up to third order in~$\delta_m$) in the bias expansion:
\begin{align}
	\delta_g(\vec{x}) =	&\ b_1\delta_m(\vec{x}) + b_{\nabla^2} R^2 \nabla^2\delta_m(\vec{x}) + b_{\delta^2} (\delta_m^2(\vec{x})-\sigma^2) + b_{s^2} (s^2(\vec{x})-\langle s^2 \rangle)	\nonumber \\
						& + b_{\Pi\Pi^{[2]}} \mathrm{Tr}[\Pi^{[1]}\Pi^{[2]}]|^{(3)}(\vec{x}) + \cdots \, ,	\label{eq:galaxy-overdensity}
\end{align}
with~$\sigma^2 = \langle \delta_m^2 \rangle$. The coefficients $b_{\O_i} = b_1, b_{\nabla^2},b_{\delta^2}, b_{s^2}, b_{\Pi\Pi^{[2]}},\cdots$, which we refer to as bias parameters, are constants that are determined by the details of galaxy formation and evolution. Put differently, we parameterize the complex physics of galaxies by these coefficients.

On large scales and for initial conditions set by single-field inflation, this is the complete list of operators for practical purposes on large scales. Because the fluctuations are necessarily adiabatic throughout cosmic history, the evolution of the universe is controlled by a single statistical quantity, which we have chosen to represent in terms of~$\Phi$ and~$\delta_m$. On smaller scales, the evolution of different components of the universe can however lead to some new types of biasing terms.\medskip

The redshift evolution of the bias parameters, particularly~$b_1(z)$, is important for projecting the sensitivity of future surveys. If (proto-)galaxies form at a high redshift and primarily evolve with the expansion of the universe, then we expect that the large-scale comoving galaxy power spectrum remains constant, $P_{gg}(k,z) \approx P_{gg}(k,z')$. In other words, the power spectrum is due to the inhomogeneities in the density field at the time they are formed, as one would find in Lagrangian biasing. This holds if
\begin{equation}
	b_1(z) = D(z)^{-1}\, b_1(z=0) \, ,	\label{eq:bias-evolution}
\end{equation}
which is a common simplifying assumption in many forecasts. Changes to the galaxy sample through mergers, for example, can however lead to behavior that differs significantly from Lagrangian biasing. When possible, we will therefore use the biases for surveys that were defined by target selection, but the assumption of this type of time evolution is also important in the context of multi-tracer analyses discussed in Section~\ref{sec:multitracer}.\medskip

\begin{figure}
	\centering
	\includegraphics{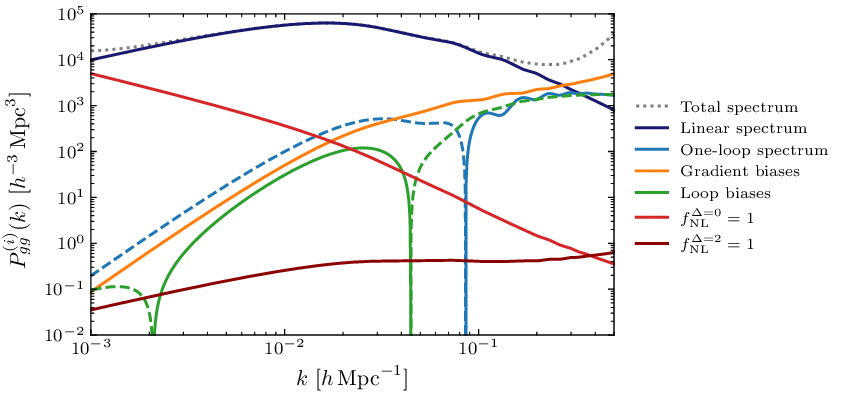}\vspace{-7pt}
	\caption{Contributions to the galaxy power spectrum~$P_{gg}^{(i)}(k)$ at $z = 0$ from the nonlinear, bias and non-Gaussian terms. The linear bias is chosen to be $b_1 = 1.6$ and the gradient biases~$b_{k^{2n}}$~(with $n \leq 2$) are taken to be unity. The bias parameters~$b_{s^2}$ and~$b_{\Pi\Pi^{[2]}}$ are calculated in the Lagrangian local-in-matter-density model~\cite{Desjacques:2016bnm}, and $b_{\delta^2}$~is calculated using the halo simulation fit of~\cite{Lazeyras:2015lgp}. For comparison, we already include the scale-dependent bias from general~PNG, which we introduce in~\textsection\ref{sec:bias}. The non-Gaussian parameter~$\fnl^\Delta$ is set to unity for both cases of $\Delta = 0$ and $\Delta = 2$. The dashed lines indicate negative contributions.}
	\label{fig:ps-contributions}
\end{figure}

Given the bias expansion for the galaxy overdensity, we can write down our model for the galaxy power spectrum after including a few additional effects. First, we work in Fourier space by transforming from real-space coordinates~$\vec{x}$ to $(k, \mu)$, where $\mu$~is the cosine between the wavevector~$\vec{k}$ and the line-of-sight direction. (Note that the composite operators~$\O_i(\x)$ are simply related to~$\tilde{\O}_i(k,\mu)$ by a single Fourier integral.) Second, there are redshift-space distortions which arise from the peculiar velocities of galaxies. We account for the Kaiser effect by adding $\Delta b(\mu, z) = f(z) \mu^2$, with the linear growth rate $f \equiv \d\hskip-1pt\log{D}/\d a$, to the linear bias~$b_1$~\cite{Kaiser:1987qv}. Collecting all terms in the bias expansion at linear order in~$\delta_m(\k)$ therefore results in
\begin{equation}
	b(k, \mu, z) = b_1(z) + f(z) \mu^2 + \sum_{n>0} b_{k^{2n}}\hskip-1pt(z) (k R_*)^{2n}\, ,	\label{eq:linear-bias}
\end{equation}
where the last term are the gradient biases in Fourier space with the comoving Lagrangian radius~$R_*$ of the halos of interest. Third, we describe gravitational nonlinearities in the very mildly nonlinear regime via the one-loop terms~$P_{22}$ and~$P_{13}$ in standard Eulerian perturbation theory. Finally, the leading-order stochastic contribution is given by~$P_{\epsilon_0}$. Putting it all together, we therefore model the theoretical anisotropic power spectrum for two galaxy \mbox{samples~$A$ and~$B$ as follows:}
\begin{align}
	P^\mathrm{th}_{AB}(k,\mu,z)	=&\ b^A(k) b^B(k) \!\left[ P_\mathrm{lin}(k) + P_{22}(k) + P_{13}(k) \right] + P_{\epsilon_0}\delta_{AB}	\nonumber	\\[1pt]
								 &+ \left[ b^A(k) b^B_{\delta^2} + b^A_{\delta^2} b^B(k) \right]\! P_{\delta^2}(k) + \left[ b^A(k) b^B_{s^2} + b^A_{s^2} b^B(k) \right]\! P_{s^2}(k)	\nonumber	\\[3pt]
								 &+ b^A_{\delta^2} b^B_{\delta^2} \!\left(P_{\delta^2\delta^2}(k) - 2\sigma^4\right) + \left(b^A_{\delta^2} b^B_{s^2} + b^A_{s^2} b^B_{\delta^2}\right)\! \left(\!P_{\delta^2 s^2}(k) - \tfrac{4}{3}\sigma^4\hskip-0.5pt\right)	\label{eq:Pg-th}	\\[4pt]
								 &+ b^A_{s^2} b^B_{s^2}\! \left(\!P_{s^2 s^2}(k) - \tfrac{8}{9}\sigma^4\hskip-0.5pt\right) + \tfrac{3}{2}\! \left[b^A(k) b^B_{\Pi\Pi^{[2]}} + b^A_{\Pi\Pi^{[2]}} b^B(k) \right]\! P_{\Pi\Pi^{[2]}}(k)\,,	\nonumber
\end{align}
where we have suppressed the dependence on~$\mu$ and~$z$ on the right-hand side, explicitly define all loop and bias contributions in Appendix~\ref{app:forecasting}, and $\delta_{AB}$ is the Kronecker delta. To illustrate these terms, we display the individual contributions in Fig.~\ref{fig:ps-contributions}.%

In practice, we compute the linear matter power spectrum~$P_\mathrm{lin}(k)$, which underlies all contributions of the theoretical galaxy power spectrum~$P^\mathrm{th}(k)$, with the Boltzmann solver~CAMB~\cite{Lewis:1999bs} using a fiducial $\Lambda$CDM~cosmological model based on the Planck~2018 TT, TE, EE + lowE + lensing + BAO best-fit cosmology~\cite{Planck:2018vyg}. In Table~\ref{tab:parameters},%
\begin{table}
	\centering
	\begin{tabular}{l l S[table-format=1.5]}
			\toprule
		Parameter 						& Description															& {Fiducial value}			\\
			\midrule[0.065em]
		$\omega_b$						& Physical density of baryons $\omega_b \equiv \Omega_b h^2$			& 0.02242					\\
		$\omega_c$						& Physical density of cold dark matter $\omega_c \equiv \Omega_c h^2$	& 0.11933					\\
		$100\theta_s$					& Angular size of the sound horizon at recombination					& 1.04119					\\
		$\tau$							& Optical depth due to reionization										& 0.0561					\\
		$\log{(\num{e10}A_\mathrm{s})}$	& Logarithm of the primordial scalar amplitude							& 3.047						\\
		$n_\mathrm{s}$					& Scalar spectral index													& 0.9665					\\
			\midrule[0.065em]
		$\fnl^\Delta$					& Non-Gaussian amplitude												& 0							\\
		$b_1$							& Linear bias															& {$b_1(z)$\hspace{12pt}}	\\
		$b_{k^{2n}}$					& Gradient biases ($n \leq 2$)											& 0							\\
		$b_{\delta^2}$					& Quadratic bias														& 0							\\
		$b_{s^2}$						& Tidal bias															& 0							\\
		$b_{\Pi\Pi^{[2]}}$				& Evolution bias														& 0							\\
			\bottomrule
	\end{tabular}
	\caption{Parameters of the fiducial $\Lambda$CDM~model, based on the Planck~2018 best-fit cosmology~\cite{Planck:2018vyg}, with the sum of neutrino masses $\sum m_\nu = \SI{0.06}{eV}$, and the biasing model employed in our forecasts. Except for the optical depth, we vary these parameters in our forecasts unless stated otherwise.}
	\label{tab:parameters}
\end{table}
we list the fiducial values of the $\Lambda$CDM, non-Gaussian and bias parameters. We set the fiducial values of the three loop biases~$b_{\delta^2}$, $b_{s^2}$ and~$b_{\Pi\Pi^{[2]}}$ to~$0$ instead of using the Lagrangian local-in-matter-density model~\cite{Desjacques:2016bnm} and fit results from halo simulations~\cite{Lazeyras:2015lgp}, as we did in Fig.~\ref{fig:ps-contributions}, since we cover a larger redshift range than the fitted bias relations were derived from. This choice should have minimal impact on the results of this paper because the forecasted constraints on the loop biases are large enough for both choices of fiducial values to be consistent with each other.

Finally, to relate the theoretical galaxy power spectrum~$P^\mathrm{th}_{AB}(k,\mu,z)$ to the observed galaxy power spectrum~$P^\mathrm{obs}_{AB}(k,\mu,z)$, we account for two observational effects that are present in any real survey: redshift errors and the Alcock-Paczynski effect. Redshift errors are observational uncertainties in the measurement of galaxy redshifts,~$\sigma_{z0}(1+z)$, which we model as an exponential suppression of the power spectrum along the line of sight,
\begin{equation}
	f^A_{\sigma_z}(k,\mu,z) = \exp\hskip-1pt\left\{-k^2\mu^2 [\sigma^A_{z0}(1+z)]^2/H^2_\mathrm{fid}(z)\right\}\, ,
\end{equation}
where~$\sigma^A_{z0}$ is the root-mean-square redshift error of sample~$A$ at $z = 0$~\cite{Seo:2003pu, Zhan:2005ki}. (We neglect this effect for spectroscopic surveys by setting $\sigma^A_{z0} \equiv 0$, i.e.~$f^A_{\sigma_z} = 1$.) The Alcock-Paczynski effect arises when the true cosmology differs from the fiducial cosmology that is used to convert the measured angular positions and redshifts of the galaxies to comoving wavenumbers in Fourier space. The effect of this mapping on the density contrast is captured by the following anisotropic rescaling of the power spectrum amplitude, the wavenumbers and the angles~\cite{Alcock:1979mp}:
\begin{equation}
	f_{AP}(z) = \frac{1}{q_\parallel q_\perp^2}\, , \qquad k'(k, \mu, z) = k \sqrt{\frac{\mu^2}{q_\parallel^2} + \frac{1-\mu^2}{q_\perp^2}}\, , \qquad \mu'(\mu, z) = \mu \hskip-1pt\Big/\hskip-1pt \sqrt{\mu^2 + (1-\mu^2) \frac{q_\parallel^2}{q_\perp^2}}\, ,
\end{equation}
where~$q_\parallel(z) \equiv H_\mathrm{fid}(z)/H(z)$ and $q_\perp(z) \equiv D_A(z)/D_A^\mathrm{fid}(z)$, with the angular diameter distance~$D_A(z)$ to redshift~$z$. To summarize, our model for the observed galaxy power spectrum in redshift space therefore is
\begin{equation}
	P^\mathrm{obs}_{AB}(k,\mu,z) = f_{AP}(z) \sqrt{f^A_{\sigma_z}(k',\mu',z)f^B_{\sigma_z}(k',\mu',z)}\, P^\mathrm{th}_{AB}(k',\mu',z) \, ,	\label{eq:Pg-obs}
\end{equation}
where $P^\mathrm{th}_{AB}(k,\mu,z)$ is given by~\eqref{eq:Pg-th}.

\subsection{Scale-Dependent Bias}
\label{sec:bias}

The conventional description of biasing applies when the initial conditions are Gaussian. Gaussian initial conditions do not correlate modes of different scales, i.e.\ the collapse of small-scale density fluctuations to form halos in different regions is determined by locally observable quantities, such as the matter density, that vary over cosmological distances due to the long-wavelength fluctuations. On the other hand, mode coupling in the initial conditions due to primordial non-Gaussianity can introduce long-range correlations that are purely related to the statistics of the small-scale fluctuations themselves. When interpreted through the bias expansion, this mode coupling may appear to be non-local~(scale-dependent) bias~\cite{Dalal:2007cu, LoVerde:2007ri, Matarrese:2008nc, Assassi:2015abs} and/or long-range stochastic bias~\cite{Baumann:2012bc}. We will focus on the former since the latter is typically a subdominant contribution to the signal.\medskip

The coupling of long and short modes is described by the squeezed or collapsed limits of a non-Gaussian correlation function. In single-field inflation, the bispectrum is constrained by the single-field consistency conditions to take the following form in the squeezed limit~\cite{Maldacena:2002vr, Creminelli:2004yq}:
\begin{equation}
	\lim_{\k_1 \to 0} \big\langle \zeta(\k_1) \zeta(\k_2) \zeta(\k_3) \big\rangle = -\left[(n_\mathrm{s}-1) + O(k_1^2/k_3^2)\right] P_\zeta(k_1) P_\zeta(k_3)\, (2 \pi)^3 \delta_D\!\left(\sum\nolimits_i \vec{k}_i\right) .
\end{equation}
Furthermore, the leading term, which corresponds to $\fnlloc =-\frac{5}{12} (n_\mathrm{s}-1)$, is unphysical~\cite{Pajer:2013ana, dePutter:2015vga}. The leading physical term is suppressed by~$O(k_1^2/k_3^2)$ and is due to the coupling of the short modes at horizon crossing to the curvature of the universe due to long-wavelength modes~\cite{Creminelli:2013cga}, which is typically proportional to the amplitude of equilateral non-Gaussianity,~$\fnleq$.

In the presence of additional fields, the short-scale power can depend on the long-wavelength values of these fields and not just derivatives of the metric fluctuations~\cite{Testa:2020hox}. For an additional massless field, the late-time Newtonian potential~$\Phi$ may be nonlinearly related to a light~(isocurvature) field~$\chi$ during inflation, $\Phi(\x) = \chi(\x) - \fnlloc \chi^2(\x)$. For a massive field, we have in the superhorizon limit
\begin{equation}
	\lim_{\k_1\to 0} \chi(\k_1,t) \approx c\frac{H}{k_1^{3/2}} \left(\frac{k_1}{a H}\right)^{\!\Delta}\, ,
\end{equation}
with a constant~$c$. If~$\chi$ alters the power spectrum of the short modes at horizon crossing, $k_2 = a H$, then we have a contribution to the bispectrum of the form
\begin{equation}
	\lim_{\k_1 \to 0} \big\langle \zeta(\k_1) \zeta(\k_2) \zeta(\k_3) \big\rangle' = \fnl^\Delta \left(\frac{k_1}{k_2} \right)^{\!\Delta} P(k_1)P(k_2)\, ,
\end{equation}
where we defined $\langle \ldots \rangle = (2\pi)^3\hskip0.5pt\delta_D(\sum_i \k_i)\,\langle \ldots \rangle'$. To have such a contribution to the bispectrum, we also need~$\chi$ to mix with~$\zeta$. Even without a mixing term, this same effect however appears in the collapsed limit of the trispectrum. For light fields, $\Delta = 3/2 - \sqrt{9/4 - m^2/H^2}$ such that $m = 0$ leads to local~PNG~($\Delta = 0$) as expected. For $m/H > 3/2$, $\Delta = \frac{3}{2} \pm \ii \nu$, with $\nu = \sqrt{m^2/H^2 - 9/4}$, is complex in which case the complex conjugate\hskip1pt\footnote{For heavy fields, there are two solutions of the long-wavelength behavior which scale as~$\Delta$ and its complex conjugate~$\Delta^\star$. We note that there is also a second solution $\Delta' = (3 - \Delta)$ for light fields which is however suppressed relative to~$\Delta$.} also contributes which results in the bispectrum being real valued, as required. More generally, it is possible to construct models with a wider range of real and complex values of~$\Delta$ than those generated by a single massive field~\cite{Green:2013rd, An:2018tcq, McAneny:2019epy, Green:2023ids}.

In single-field inflation, physical mode coupling must involve derivatives of the metric fluctuations~\cite{Pajer:2013ana}, whose leading behavior in the soft limit corresponds to $\Delta = 2$~\cite{Creminelli:2012ed}. One might naturally be surprised to find that the $m^2 \to \infty$ limit of~$\Delta$ for a massive field does not yield this value of single-field inflation. It was nicely explained in~\cite{Arkani-Hamed:2015bza} that $\Delta \approx 3/2 + \ii m/H$ is precisely what is expected from the single-particle wavefunction of a massive particle in an expanding universe in the high-mass limit. This contribution to the squeezed limit arises from the physical production of these particles. This particle production is Boltzmann suppressed in some simple models~\cite{Noumi:2012vr}, $\fnl^\Delta \propto \ee^{-\pi m/H}$, but can be enhanced depending on the nature of the interaction~\cite{Flauger:2016idt}. In addition, very massive fields will also produce a sub-leading contribution to the squeezed limit with $\Delta = 2$. This contribution arises from the virtual exchange of a massive field and is equivalent to a purely local interaction~(i.e.\ to integrating out the massive field). It may be surprising that a local term will produce non-local scale-dependent bias. When $T(k) \to 1$, \mbox{$\Delta = 2$ is} indeed a local term and it was long thought that~$\fnleq$ could not be measured via scale-dependent bias for this reason~(see e.g.~\cite{Assassi:2015abs}). However, because of the evolution of matter after horizon entry~($T(k) \neq 1$), local interactions during inflation are distinguishable from local processes in structure formation at late-times~\cite{Gleyzes:2016tdh}, i.e.\ equilateral~PNG can be measured in the power spectrum via scale-dependent~bias.\medskip

In the conventional picture of galaxy biasing~(see~\textsection\ref{sec:spectrum}), the density contrast of galaxies is determined by a long list of composite operators in the initial density field. The origin of the unique signal of~PNG in galaxies is that the behavior of these composite operators becomes a proxy for the light fields during inflation. For example, consider the impact of structure formation from the local amplitudes of fluctuations on a scale~$R$,
\begin{equation}
	\sigma_R(\x) = \int\! \d^3 k\, \ee^{-\ii \k \cdot \x} \int\! \frac{\d^3 p}{(2\pi)^3} \zeta(\p) \zeta(\k-\p) F(p R) \, ,
\end{equation}
for some filter function~$F(x)$. This is the local variance of the primordial fluctuations which varies from place to place. Importantly, this variance is correlated with the long-wavelength metric perturbations as follows:
\begin{equation}
	\langle \sigma_R(\k) \zeta(\k') \rangle' \propto \fnl^\Delta (k R)^\Delta P_\zeta(k) \quad \Rightarrow\ \quad \langle \sigma(\k) \delta_m(\k') \rangle' \propto \fnl^\Delta \frac{(k R)^\Delta}{k^2 T(k)} P_m(k) \, .
\end{equation}
Since the number density of galaxies at a location~$\x$ is related to~$\sigma_R(\x)$, this introduces a long-distance correlation between~$\delta_g(\x)$ and~$\delta_m$ that is non-local. The resulting non-local modification of the bias~\eqref{eq:linear-bias} is given by~\cite{Baumann:2012bc}\hskip0.5pt\footnote{It is important to note that, in principle, the scale-dependent bias is distinguishable from the expansion in~$(k R_\star)$ even for $\Delta = 2n$ for any positive integer~$n$. At $k \ll k_\mathrm{eq}$, $\T(k) \to \mathrm{const}$ and the scale-dependent bias scales as~$k^{2n-2}$ which is degenerate with~$b_1$ and~$b_{k^{2(n-1)}}$. However, since $k_\mathrm{eq} R_\star \ll 1$, the transfer function introduces a non-trivial scale dependence in the regime $k_\mathrm{eq} \lesssim k \ll R_\star^{-1}$ which is not captured in the local bias expansion.}
\begin{equation}
	b(k, \mu, z) = b_1(z) + f(z) \mu^2 + \sum_{n>0} b_{k^{2n}} (k R_*)^{2n} + A \fnl \frac{b_\phi(z)}{k^2\T(k,z)} (k R_*)^\Delta\, ,	\label{eq:scale-dependent_bias}
\end{equation}
where~$b_\phi$ is the non-Gaussian bias parameter and we took $R = R_*$. The constant~$A$ normalizes the non-Gaussian parameter~$\fnl^\Delta$ in the conventional way for~PNG in the squeezed limit of the bispectrum,\footnote{Note that for the special bispectrum configurations, when expressed in terms of the general exponent~$\Delta$, the equivalent~$\fnl^\Delta$ definitions are related by $\fnlloc = 3\fnl^{\Delta = 0}$ and $\fnleq = \fnl^{\Delta = 2}$.} so that the last term in~\eqref{eq:scale-dependent_bias} is~\cite{Matarrese:2008nc, Schmidt:2010gw, Desjacques:2011jb, Desjacques:2011mq, Giannantonio:2011ya}
\begin{align}
	b_\mathrm{NG}^\mathrm{loc}(k,z)	&= \fnlloc \frac{b_{\phi}(z)}{k^2\T(k,z)}						&\text{(local)},								\\
	b_\mathrm{NG}^\mathrm{eq}(k,z)	&= 3\fnleq \frac{b_{\phi}(z)}{k^2\T(k,z)}(kR_*)^2				&\text{(equilateral)},							\\
	b_\mathrm{NG}^\Delta(k,z)		&= 3\fnl^{\Delta} \frac{b_{\phi}(z)}{k^2\T(k,z)}(kR_*)^\Delta	&\text{(general~exponent~$\Delta \in [0, 2]$)}.	\label{eq:bias_delta}
\end{align}
Note that we do not include the physical orthogonal shape, as it is produced in single-field inflation and, therefore, has the same squeezed limit as the equilateral shape, i.e.\ corresponds to~$\Delta = 2$. Previous work on the scale-dependent bias used an orthogonal template which has $\Delta = 1$~(e.g.~\cite{Norena:2012yi}), which is however unphysical and has not been used in recent LSS~analyses.

Without any additional theoretical input, $b_\phi(z)$ and~$\fnl^\Delta$ are both unknown quantities and are completely degenerate. If that was the case, we could only measure the combination~$\fnl^\Delta b_\phi(z)$ in each redshift bin which would make it challenging to infer limits on~$\fnl^\Delta$ alone. More optimistically, we can fix~$b_\phi(z)$ in terms of the linear bias~$b_1(z)$ using the universality relations found in phenomenological models~\cite{Desjacques:2016bnm},
\begin{equation}
	b_\phi(z) = 2\delta_c (b_1(z) - p) \, ,	\label{eq:universality}
\end{equation}
where $\delta_c = 1.686$ is the critical overdensity for spherical collapse and $p = 1$ if we assume a universal halo mass function. We will adopt this relation in our forecasts, but refer to e.g.~\cite{Barreira:2020ekm, Barreira:2021ueb, Barreira:2021ukk, Barreira:2022sey} for a recent discussion of the validity and risks of this choice.\medskip

To conclude, we note that we incorporate the general scale-dependent bias~\eqref{eq:scale-dependent_bias} in the galaxy power spectrum as described in~\eqref{eq:Pg-th} and~\eqref{eq:Pg-obs}, i.e.\ $b^A = b^A_\Delta(k, \mu, z)$. In the rest of the paper, we follow~\cite{Gleyzes:2016tdh} in assuming a minimum halo mass of $\num{e13} M_\odot$, which corresponds to $R_* \approx \SI{2.66}{\MpcPerh}$, and truncating the non-local gradient bias expansion at $n = 2$. We illustrate the effect of~$b_\mathrm{NG}^\Delta$ on the linear galaxy power spectrum in Fig.~\ref{fig:bias} and compare these non-Gaussian contributions to the nonlinear galaxy power spectrum in Fig.~\ref{fig:ps-contributions} at redshift $z=0$ for linear bias $b_1 = 1.6$. We see from the latter figure that the $\Delta = 0$ contribution does not mimic the behavior of any other bias or nonlinear terms at low~$k$. On the other hand, the $\Delta = 2$ contribution is both much smaller at $k < \knl$ and similar in behavior to those Gaussian contributions.

\section{Information Content of the Galaxy Power Spectrum}
\label{sec:forecasts}

Our goal in this section is to understand how to design a galaxy survey to best measure~$\fnl^\Delta$ for $\Delta \in [0, 2]$. To do so, we will explore the Fisher forecasts for~$\fnl^\Delta$ as a function of~$\Delta$ for a variety of survey configurations and biasing models.\footnote{In Appendix~\ref{app:scaling}, we also discuss our ability to measure the scaling behavior of the non-Gaussian bias, where we also include~$\Delta$ as an additional free parameter in the Fisher matrix.} We will compare these results with analytic estimates to understand what aspects of the surveys drive major improvements in sensitivity. While our forecasts will be rooted in specific survey configurations, our goal is to isolate the features of these surveys that impact the forecasted constraining power the most.

\subsection{Fisher Information Matrix}
\label{sec:fisher}

In the following, we describe the well-known Fisher matrix forecasting techniques for the galaxy power spectrum that we use. We review the basics here and refer to Appendix~\ref{app:forecasting} for the experimental specifications of the surveys.\medskip

The Fisher matrix for a set of galaxy samples, labeled by $\{A, B, \ldots\}$, is defined by
\begin{equation}
	F_{\alpha\beta} = \sum_{z_i} V(z_i) \int_{-1}^{1}\! \frac{\d\mu}{2} \int_{\kmin}^{\kmax} \frac{\d k\,k^2}{2\pi^2}\, \frac{1}{2} \mathrm{Tr}\left[ \mathbf{C}_{,\alpha} \mathbf{C}^{-1} \mathbf{C}_{,\beta} \mathbf{C}^{-1} \right] ,	\label{eq:fisher}
\end{equation}
where $\theta_{\alpha, \beta}$ are the model parameters, the trace is over the samples and we defined $\mathbf{C}_{,\alpha} \equiv \partial\mathbf{C}/\partial\theta_\alpha$. The integrals are discretized, with the integration over wavenumbers~$k$ being limited by the minimum wavenumber~$\kmin$ and the maximum wavenumber~$\kmax$, and bins of width $\Delta k = \kmin$. We note that we vary all parameters listed in Table~\ref{tab:parameters} throughout this paper, unless specified otherwise. The elements of the matrix~$\mathbf{C}$ are given by
\begin{equation}
	\mathbf{C}_{AB}(k,\mu,z) \equiv P^\mathrm{obs}_{AB}(k,\mu,z) + \delta_{AB} N_A(k,\mu,z) \, ,
\end{equation}
with the shot noise term $N_{A}(k,\mu,z) = \bar{n}_A(z)^{-1}$, where $\bar{n}_A(z)$ is the average comoving number density of tracers at redshift~$z$. Shot noise arises because we observe a finite number of discrete objects in galaxy surveys. (Note that we absorbed the stochastic term~$P_{\epsilon0}$ of~\eqref{eq:Pg-th} into~$N_{A}$.) We treat the redshift bins as spherical shells so that the volume of the $i$-th redshift bin is
\begin{equation}
	V(z_i) = \frac{4\pi}{3} \fsky \left[ d_c(z_i + \Delta z_i/2)^3 - d_c(z_i - \Delta z_i/2)^3 \right] ,
\end{equation}
where $\fsky$~is the sky fraction observed by the survey, $d_c(z)$~is the comoving distance to redshift~$z$ and $\Delta z_i$~is the width of the $i^\mathrm{th}$~redshift bin.\medskip

To ensure that only modes under perturbative control in Eulerian perturbation theory and the bias expansion enter our Fisher matrix calculation, and to exclude long-wavelength modes inaccessible to a given survey due to its finite volume, we restrict the integral in~\eqref{eq:fisher} to a maximum wavenumber~$\kmax$ and a minimum wavenumber~$\kmin$, respectively.\footnote{Alternatively, we could have included the correlations between $k$~bins for small wavenumbers and the theoretical uncertainties following~\cite{Baldauf:2016sjb, Chudaykin:2020hbf, Moreira:2021imm} for large wavenumbers. Our approximate and conservative treatment using $k \in [\kmin, \kmax]$ follows the common procedure in Fisher forecasts.} We make a conservative choice for~$\kmax$ by taking the smaller of the two scales indicating the range of validity of perturbation theory and the bias expansion. To be precise, we take it to be the minimum of the nonlinear scale\hskip1pt\footnote{We use $\knl = \frac{\pi}{2 R_\mathrm{NL}}$, with the radius~$R_\mathrm{NL}$ at which the variance of linear fluctuations is $\sigma_R = 1/2$ at a given~$z$.}~$\knl$ and the wavenumber~$k_\mathrm{halo} \approx \SI{0.19}{\hPerMpc}$ associated with the Lagrangian scale corresponding to the minimum halo mass of the surveyed population which we assume to be $R_* \approx \SI{2.66}{\MpcPerh}$~(cf.~\textsection\ref{sec:bias}), i.e.\ $\kmax(z) = \min\{\knl(z), k_\mathrm{halo}\}$. For~$\kmin$, we assume a spherical survey geometry per redshift bin which results in a conservative minimum wavenumber $k_{\mathrm{min},i} = 2\pi\hskip1pt[ 3 V(z_i) / (4\pi) ]^{-1/3}$.\medskip

For most of the forecasts in Sections~\ref{sec:forecasts} and~\ref{sec:multitracer}, we use a futuristic spectroscopic survey with a billion objects divided into 10~redshift bins with $z \leq 5$, which we will hereafter refer to as the \textit{billion-object survey}. We split the galaxy sample of this survey into two populations based on their linear bias at $z = 0$, chosen as $b_1^{(1)}(z=0) = 2.0$ and $b_1^{(2)}(z=0) = 1.2$. We model their redshift evolution according to~\eqref{eq:bias-evolution} as previously discussed in~\textsection\ref{sec:spectrum}. In addition, we assume a constant number density across all redshift bins and both galaxy samples, $\bar{n}^{(1)}_g(z) = \bar{n}^{(2)}_g(z) = \mathrm{const}$. The details of this and other surveys used in our forecasts, i.e.\ the Baryon Oscillation Spectroscopic Survey~(BOSS)~\cite{BOSS:2016wmc}, the Dark Energy Spectroscopic Instrument~(DESI)~\cite{DESI:2016fyo}, Euclid~\cite{EUCLID:2011zbd, Euclid:2021icp}, Vera~Rubin Observatory's~LSST~\cite{LSSTScience:2009jmu}, the Spectro-Photometer for the History of the Universe, Epoch of Reionization and Ices Explorer~(SPHEREx)~\cite{SPHEREx:2014bgr}, MegaMapper~\cite{Ferraro:2019uce, Schlegel:2022vrv} and the billion-object survey are provided in Appendix~\ref{app:forecasting}. Except for~SPHEREx and the billion-object survey, we treat all surveys as single-tracer surveys by combining different groups of tracers~(if available) into a single effective number density and bias. The impact of this should not significantly impact our forecasts as we explain in Appendix~\ref{app:forecasting}.\medskip

Unless mentioned otherwise, we marginalize over all bias parameters separately for each sample and redshift bin, and include CMB~information on the $\Lambda$CDM~parameters~(but not on~PNG) from~Planck throughout the paper.\footnote{When marginalizing over the full biasing model for the billion-object survey, for instance, the difference between using the $\Lambda$CDM~covariance from Planck or CMB-S4~\cite{Abazajian:2019eic, CMB-S4:2022ght} for $\Delta \lesssim 1.5$ is less than two percent. In fact, this CMB~information decreases~$\sigma(\fnl^\Delta)$ only by a few percent for $\Delta \lesssim 1.3$, and by a maximum of~20\%~(15\%) with~Planck~(CMB-S4) at $\Delta = 2$.} We achieve this by computing the $\Lambda$CDM~Fisher matrix directly following the Fisher methodology and the experimental specifications of~\cite{Baumann:2017gkg}~(see also~\cite{Allison:2015qca} for the latter), marginalized over the optical depth~$\tau$. We then combine this marginalized Planck~Fisher matrix with the respective LSS~Fisher matrix for the five $\Lambda$CDM~parameters, the non-Gaussian parameter~$\fnl^\Delta$ and the galaxy bias parameters~$b_{\O_i}$. We treat each redshift bin with mean redshift~$\bar{z}_i$ as an independent survey and independently marginalize over the biases within each redshift bin.

\subsection{Information and Survey Design}
\label{sec:information}

Using Fisher forecasts, our aim is to understand the possible strategies for optimizing the measurement of~$\fnl^\Delta$ for shapes including but not limited to local~PNG. The parameters that we most directly control in the design of a survey are the sky fraction~($\fsky$), the redshift range~($z\in[\zmin,\zmax]$) and the number density of tracers~[$\bar{n}_g(z)$]. However, the actual statistical power of the survey is controlled by the smallest~($\kmin$) and largest~($\kmax$) wavenumbers that can be reliably measured, the linear bias of the sample~[$b_1(z)$], etc. These factors are influenced by the details of the survey, but can also be affected by systematics and astrophysics.\medskip

\begin{figure}
	\centering
	\includegraphics{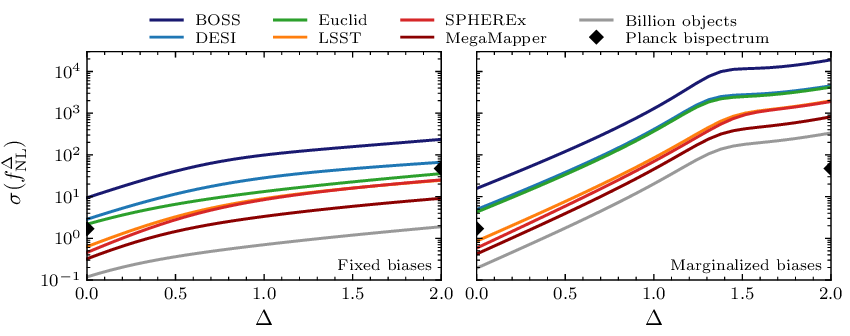}\vspace{-7pt}
	\caption{Forecasted constraints on~$\fnl^\Delta$ for general scaling $\Delta \in [0, 2]$ of the scale-dependent bias, for current and future surveys, compared to the Planck~bispectrum constraints for local and equilateral non-Gaussianity. We also marginalize over the five $\Lambda$CDM~parameters, use Planck~results as priors for the $\Lambda$CDM~parameters, and either fix the bias parameters~(\textit{left}) or marginalize over them~(\textit{right}). Note that the SPHEREx~forecast uses the five-tracer sample and the billion-object survey forecast uses the double-tracer sample as provided in Appendix~\ref{app:forecasting}. For all the other surveys, we use their combined single-tracer sample.}
	\label{fig:future_surveys}
\end{figure}

We show the overall landscape of future power spectrum measurements of~$\fnl^\Delta$ for $\Delta \in [0, 2]$ in~Fig.~\ref{fig:future_surveys}. %
Qualitatively, the forecasts show the overall behavior that we might expect: (i)~constraints at $\Delta = 0$ are generally much stronger than at $\Delta = 2$, (ii)~higher-order biasing affects larger~$\Delta$ more than smaller~$\Delta$, (iii)~larger surveys have more constraining power, and (iv)~future LSS~surveys will be able to improve over Planck bispectrum constraints from the galaxy power spectrum for $\Delta = 0$, but not for $\Delta = 2$ without additional information on bias parameters. In detail, the forecasts however have elements that are harder to understand without further investigation. First, the forecasts show clear features at values of $\Delta \sim 1.4$ that depend on the marginalization over biasing parameters~(see also Fig.~\ref{fig:kmax} below). This suggests a qualitative change in where the constraining power is coming from as we vary~$\Delta$. Second, we see that~SPHEREx and~LSST produce very similar forecasts despite having very different strengths and weaknesses. Furthermore, both significantly exceed the constraining power of Euclid and~DESI which have the advantage of being three-dimensional~(spectroscopic) surveys. In order to make sense of these forecasts, we will break them down according to the number density~(noise), scales~($\kmin$/$\kmax$), the fiducial bias~[$b_1(z)$], the survey geometry~\mbox{($\fsky$, $\zmax$)} and redshift errors~($\sigma_{z0}$) in the following.

\subsubsection*{Dependence on Number Density}

The most basic parameter in any experiment is the signal to noise. In a galaxy survey, shot noise is the dominant~(irreducible) noise source. At fixed volume, increasing the number of objects increases the range of wavenumbers~$k$ where~$P_{gg}(k)$ is measured with at least signal to noise of order one, $P_{gg}(k)/N(k) \gtrsim O(1)$.\medskip

For most cosmological parameters, sample variance presents an additional irreducible source of noise. However, for scale-dependent bias, the parameter~$\fnl^\Delta$ can, in principle, be measured without sample variance when the signal to noise is large for multiple tracers~\cite{Seljak:2008xr}. Concretely, the relationship between~$\delta_g(\k)$ and~$\delta_m(\k)$ in~\eqref{eq:galaxy-overdensity} is completely deterministic and is therefore not limited by sample variance. However, this only works if we can measure the same Fourier mode with high signal to noise for tracers with different values of~$b_\phi$. The noise introduced by sample variance is proportional to~$P(k)$, while shot noise is independent of~$k$. As a result, sample-variance cancellation dramatically alters how different scales contribute to the measurement of~$\fnl^\Delta$ and, therefore, the qualitative understanding of our forecasts. This is why it is important to distinguish from the outset to what degree multi-tracer sample-variance cancellation \mbox{is relevant in our forecasts.}\medskip

\begin{figure}
	\centering
	\includegraphics{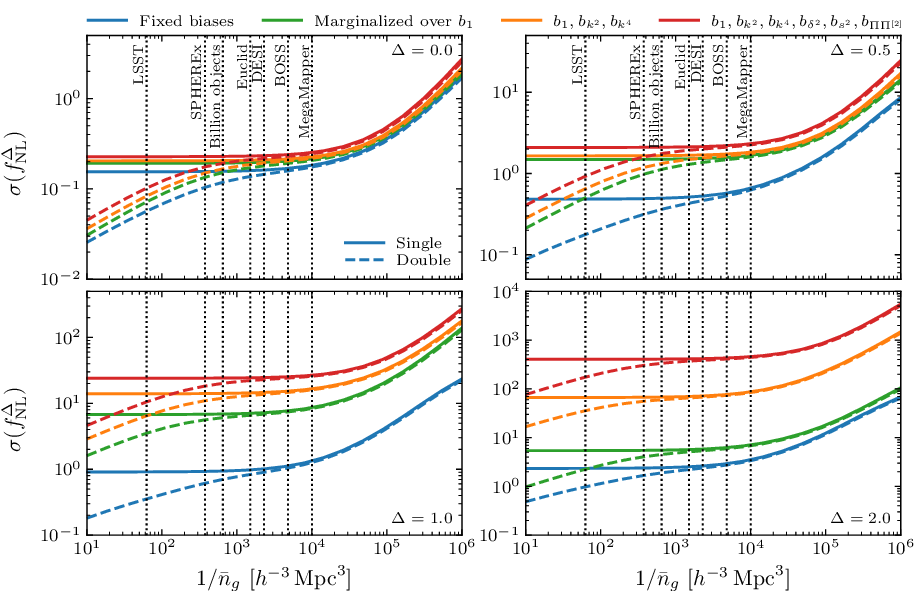}\vspace{-7pt}
	\caption{Dependence of~$\sigma(\fnl^\Delta)$ on the shot noise, i.e.\ the inverse survey number density~$\bar{n}_g^{-1}$, for the billion-object survey for four representative values of~$\Delta$. The solid lines show the single-tracer forecasts and the dashed lines display the double-tracer results. The different colors indicate which bias parameters are marginalized when calculating~$\sigma(\fnl^\Delta)$. The vertical lines mark the averaged shot noise of various surveys which are calculated by taking the average of the number densities of the middle~50\% of redshift bins of each survey, i.e.\ they should be taken as an approximate illustration of the noise level in these surveys.}
	\label{fig:nbar}
\end{figure}

The impact of shot noise in both single- and multi-tracer scenarios is illustrated in Fig.~\ref{fig:nbar}.%
Two points should be very clear from these curves: (1)~sample-variance cancellation has the potential to dramatically improve the sensitivity to~$\fnl^\Delta$ for all values of~$\Delta$, and (2)~realistic current and future surveys are very far from being in the regime where sample-variance cancellation has a large impact. In fact, most near-future surveys are, at best, at the boundary between single and multiple tracers.\footnote{Note that the curves in Fig.~\ref{fig:nbar} are computed using the spectroscopic billion-object survey~(cf.~Table~\ref{tab:billion-specs}). This means that the photometric~LSST is not a direct comparison and the line labeled as~LSST only actually indicates a spectroscopic follow-up of~LSST. The potential sensitivity in a multi-tracer analysis of~LSST itself will be discussed in~\textsection\ref{sec:LSST_multi}.} As a result, sample-variance cancellation is important for highly accurate forecasts, but the qualitative behavior is consistent with single-tracer forecasts. We therefore use a single-tracer emphasis for the rest of this section. We will revisit the advantages offered by multi-tracer techniques and how astrophysics affects the constraints on~$\fnl^\Delta$ in Section~\ref{sec:multitracer}.

\subsubsection*{Dependence on Scales}

The defining feature of scale-dependent bias is that it is enhanced at small wavenumbers/large distances. For $\Delta < 2$ and $\fnl^\Delta \neq 0$, we have $P_{gg}(k) \gg b_1^2 P_m(k)$ as $k \to 0$. Since the noise is also a function of~$k$, it is however not a given that the information resides at small~$k$ for all $\Delta < 2$. We now investigate this analytically in the Fisher matrix and numerically in our forecasts.\medskip

For a single-tracer analysis where we know the bias parameters exactly, i.e.\ $b_{\O_i}$ are held fixed, the Fisher information for a fiducial $\fnl^\Delta = 0$ is given by
\begin{equation}
	F_{\fnl^\Delta \fnl^\Delta} = \sum_{z_i} V(z_i) \int\! \frac{\d^3k}{(2\pi)^3} \frac{[6 b_1 b_\phi(z_i) (k R_*)^\Delta]^2 P_m(k,\mu,z_i)^2}{2 k^4\T(k,z_i)^2\, [b_1^2 P_m(k,\mu,z_i) + N(k,\mu,z_i)]^2} \, .
\end{equation}
In the high signal-to-noise regime, $b_1^2 P_m(k) \gg N$, this becomes
\begin{equation}
	F_{\fnl^\Delta \fnl^\Delta} \approx \frac{9}{\pi^2} \sum_{z_i} \frac{b_\phi^2(z_i)}{b_1^2(z_i)} R_*^{2\Delta} V(z_i) \int\! \d k\, \frac{k^{2\Delta - 2}}{\T(k,z_i)^2}\, .	\label{eq:Delta_transition}
\end{equation}
For $2\Delta < 1$, the integral over~$k$ is dominated by~$\kmin$ and the Fisher information arises from the smallest scales. However, due to the limits placed by noise and volume, realistic surveys are far from the asymptotic $k \to 0$ regime. For modes with the largest signal to noise, $k \approx k_\mathrm{eq} \approx \SI{e-2}{\hPerMpc}$, the matter power spectrum is flat, $P_m(k \sim k_\mathrm{eq}) \propto k^0$ and, therefore, $\T(k,z)^2 \sim k^{-1}$. In this regime, repeating our analysis of~\eqref{eq:Delta_transition} shows that the integral is dominated by large~$k$ for $\Delta \gtrsim 0$. In this regard, we expect that the transition from low to high~$k$ is gradual as we change~$\Delta$.\medskip\vspace{-2pt}

\begin{figure}
	\centering
	\includegraphics{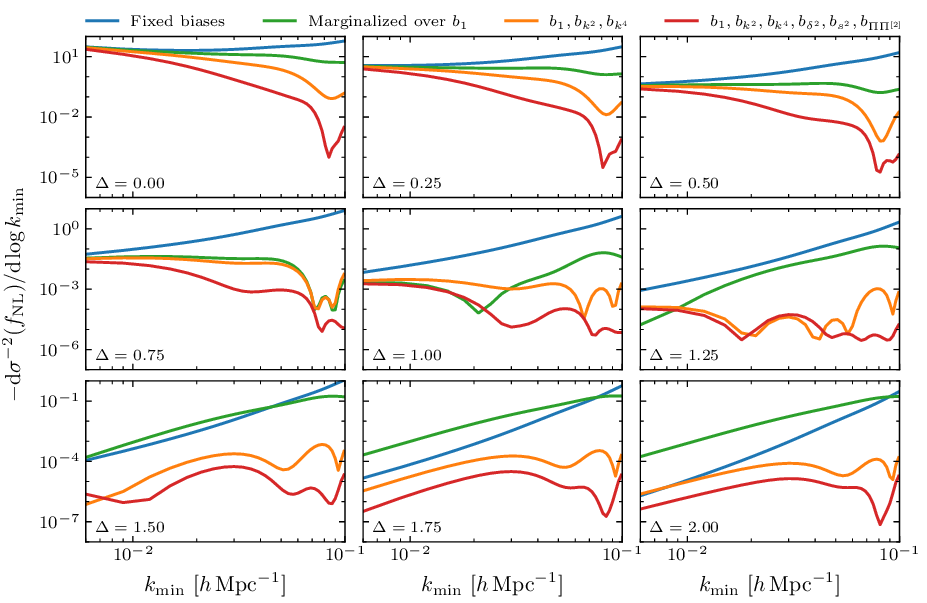}\vspace{-8pt}
	\caption{The information density with respect to~$\log\kmin$ in the billion-object survey, with fixed $\Lambda$CDM~cosmology, and the same~$\kmin$, $\Delta k = \SI{0.003}{\hPerMpc}$ and $\kmax = \SI{0.2}{\hPerMpc}$ in each redshift bin. The different colors indicate which bias parameters are marginalized when calculating~$\sigma(\fnl^\Delta)$ using the double-tracer version of the survey. A negative~(positive) slope of the curves indicates that there is more~(less) information on smaller wavenumbers~$k$. We observe that the information density is dominated by small~(large) wavenumbers \mbox{for small~(large) scaling exponents.}}\vspace{-5pt}
	\label{fig:kmin}
\end{figure}

This behavior is seen in forecasts by varying~$\kmin$ as illustrated in Fig.~\ref{fig:kmin}.%
Since increasing the minimum wavenumber~$\kmin$ removes information, the constraints~$\sigma(\fnl^\Delta)$ always decrease. We therefore show~$\d\sigma^{-2}(\fnl^\Delta)/ \d\hskip-1pt \log\kmin$ because this quantity provides a more quantitative illustration of how much information is being lost as we change~$\kmin$.\footnote{We consider the logarithmic derivative in order to numerically extract the relevant scaling that we analytically discussed in the previous paragraph. We additionally note that $\log\kmin$ is implicitly normalized by~\SI{1}{\hPerMpc}.} For $\Delta = 0$, the largest impact is for the smallest values of~$\kmin$, indicating that the lowest wavenumber has the most information. As we increase $\Delta > 0$, we see that this trend changes: the curves are mostly flat at small~$\kmin$ for $\Delta \lesssim 0.5$. This is a reflection of the fact that there is no sharp transition in~$\Delta$ due to the impact of~$\T(k,z)$ in~\eqref{eq:Delta_transition}.

As we move from large to small scales~(or small to large~$k$), we are increasingly sensitive to the marginalization over the bias parameters. The information density in Fig.~\ref{fig:kmin} at large wavenumbers becomes highly suppressed as we marginalize so that the true constraining power remains at small wavenumbers. The oscillatory behavior in these figures highlights that the baryon acoustic oscillations contained in the transfer function are imprinted in the non-Gaussian signal and are not absorbed into the bias expansion.\medskip

\begin{figure}
	\centering
	\includegraphics{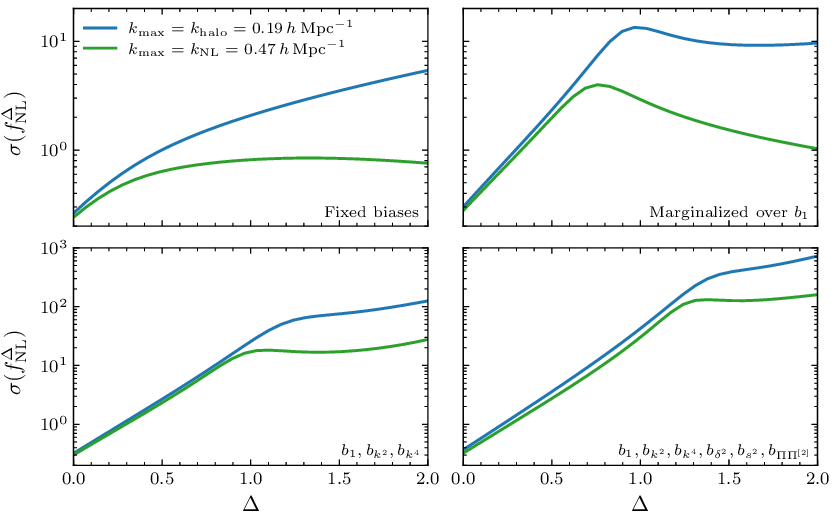}\vspace{-5pt}
	\caption{Dependence of $\sigma(\fnl^\Delta)$ on the maximum wavenumber where we take $\kmax = k_\mathrm{halo}$ and $\kmax = \knl(z)$, with the latter being less conservative at higher redshifts. Here, we combine the middle two redshift bins~($2 \leq z \leq 3$) of the double-tracer billion-object survey into a single comoving box with a total of about \num{1.3e8}~objects. We show this dependence of the constraining power for four biasing models. The choice of~$\kmax$ barely impacts~$\sigma(\fnl^\Delta)$ at small~$\Delta$, but becomes significant for large~$\Delta$. We explain the origin of the feature around $\Delta = 1$ in the main text.}
	\label{fig:kmax}
\end{figure}

At the same time, this short-distance information remains however crucial for breaking degeneracies between bias and cosmological parameters. This is shown in Fig.~\ref{fig:kmax},%
which illustrates how the maximum wavenumber~$\kmax$ impacts the constraining power. We see that the constraint on~$\fnl^\Delta$ improves by roughly a factor of five to nine~(depending on the biasing model) for $\Delta = 2$ when we choose the less conservative $\kmax = \knl$. This improvement is consistent with the number of additional modes available between~$k_\mathrm{halo}$ and~$\knl$, and the fact that our signal depends on wavenumbers. This difference becomes even more pronounced at higher redshifts since $\knl(z)$~increases significantly, while~$k_\mathrm{halo}$ is independent of redshifts. For all of our other forecasts, we however use the more conservative choice of $\kmax(z) = \min\{k_\mathrm{halo}, \knl(z)\}$ as noted in~\textsection\ref{sec:fisher}.

When marginalizing over~$b_1$, we notice a feature around $\Delta \approx 1$ in Fig.~\ref{fig:kmax} and not around $\Delta = 0.5$ as we might expect from~\eqref{eq:Delta_transition}. This is a consequence of the information on~$\fnl^\Delta$ moving to larger~$k$ where $T(k) \neq 1$. Specifically, if we assume $P_m(k) \propto k^n$ for large~$k$, then $k^2 T(k) \propto k^{n/2+3/2}$, which implies
\begin{equation}
	P_{gg}(k) \approx \left(b_1 + 2 b_\phi \fnl^\Delta C (k R_*)^{\Delta-\frac{n}{2}-\frac{3}{2}} + \ldots\right)^{\!2} P_m(k) \, ,	\label{eq:Pg-approx}
\end{equation}
for some constant~$C$. Using $n \approx -1$ for $k \sim \SI{0.1}{\hPerMpc}$~\cite{Orban:2011rx}, we notice that $\Delta - \frac{3}{2} - \frac{n}{2} \approx \Delta - 1$. At large wavenumbers, the ``scale-dependent bias'' for $\Delta \approx 1$ is therefore degenerate with the linear bias~$b_1$, resulting in the observed feature.

\subsubsection*{Dependence on Fiducial Biases}

All else being equal, a single-tracer survey with the largest possible (absolute) value of $b_\phi \propto (b_1 -1)$ will yield the most sensitive measurement of~$\fnl$. It may therefore seem self-evident that selecting highly biased targets is a central tool in the search for~PNG. Having said that, in practice, target selection involves numerous factors that lie beyond the scope of these simple forecasts. Yet, when it comes to understanding the performance of a given survey, the biases of the objects in their sample will strongly influence the overall sensitivity. As a result, it is important to separately understand the role that the fiducial biases play from the aspects of the survey that we can control more directly.\medskip

\begin{figure}
	\centering
	\includegraphics{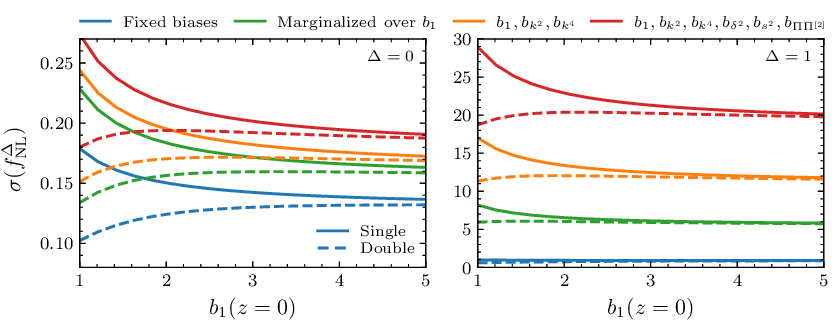}\vspace{-5pt}
	\caption{Dependence of~$\sigma(\fnl^\Delta)$ on the fiducial value of the linear bias for both the single-tracer~(solid) and double-tracer~(dashed) versions of the billion-object survey for $\Delta = 0$~(\textit{left}) and $\Delta = 1$~(\textit{right}). We show this dependence as a function of the linear bias~$b_1(z=0)$ of the single-tracer survey in which case we see the improvement in the constraining power of the survey for larger~$b_1$. In the double-tracer scenario, for which the galaxy sample is split into two populations of equal number density with $b_1^{(A,B)}(z=0) = b_{1}(z=0) \pm 0.4$, the effects of sample-variance cancellation can be observed. (Note that the $y$-axis is on a linear and not on a logarithmic scale like in the previous figures.)}
	\label{fig:b1}
\end{figure}

For a single-tracer analysis, we only measure $b_\phi \fnl^\Delta$. At fixed redshift and assuming $b_\phi \propto (b_1-1)$ from~\eqref{eq:universality}, we therefore have
\begin{equation}
	\sigma(\fnl^\Delta)_\mathrm{single}^\text{fixed-$z$} \propto \frac{1}{b_1 - 1} \, .
\end{equation}
As a result, we can expect a large improvement in a single-tracer analysis from choosing a sample with $b_1 \gtrsim 2$. Figure~\ref{fig:b1}%
shows how the value of $b_1(z=0)$ influences the overall constraining power of our billion-object survey in both single- and multi-tracer analyses. We see that~$\sigma(\fnl)$ increases sharply when the bias approaches one. It does not diverge because of the redshift dependence of~$b_1(z)$ which ensures that some of the redshift bins have $b_\phi \neq 0$. On the other hand, sample-variance cancellation in the multi-tracer configuration results in a much weaker dependence of~$\sigma(\fnl^\Delta)$ on the linear bias and, in fact, benefits from having one of the tracers with $b_1^B(z=0) < 1$. At the same time, we however see that taking larger values of~$b_1(z)$ in a single-tracer analysis closes the gap to the double-tracer case for $b_1(z=0) \gtrsim 3$.\medskip

Technically speaking, we cannot choose the linear bias~$b_1(z)$ and the number densities~$\bar{n}_g(z)$ independently, in particular since more highly biased objects are more rare. In practice, traditional spectroscopic surveys are typically limited by the number of spectra that can be measured and not the number of objects in the universe. However, other types of surveying techniques, such as line intensity mapping~(see e.g.~\cite{Camera:2014bwa, Li:2017jnt, MoradinezhadDizgah:2018lac, CosmicVisions21cm:2018rfq, PUMA:2019jwd, Karagiannis:2020dpq} in the context of~PNG), may get closer to these fundamental limits. We refer to~\cite{Gleyzes:2016tdh} for a discussion of the sensitivities of surveys in the regime where the galaxy power spectrum analysis is limited by the number of available objects.

\subsubsection*{Dependence on the Volume and Redshift}

Given a single redshift bin with volume~$V$ and a fixed choice of targets, it is easy to see the impact of this choice on~$\kmin$ and~$\kmax$. However, when we design a survey, the choice of redshift range, volume and targets is not fixed, but is part of the survey design. The redshift range and volume affect~$\kmin$ and~$\kmax$, but they also change the biases of the targets and the number densities of objects in each redshift bin. For the purpose of designing a survey, we therefore want to understand the optimal choices for the sky fraction~$\fsky$ and the redshift range as given by~$\zmax$.\medskip

It is important to compare survey strategies assuming ``constant effort''. In practice, it is the time for acquiring spectra that limits the sensitivity of a spectroscopic survey. This is why we vary~$\fsky$ for a given redshift range while holding the total number of objects fixed. As a result, there is a trade-off between increasing the volume and increasing the shot noise, which needs to be optimized. When it comes to varying the redshift range, the time required to observe each object changes with redshift which means that the total number of objects in the sample decreases with larger~$\zmax$. To hold the observational effort fixed, we use a simplified model where it takes twice the amount of time to obtain spectra for galaxies at $z > 3$ than for galaxies at $z < 3$~(cf.~\cite{DESI:2022lza}). We further assume that it is desirable to maintain a constant level of shot noise over the survey volume, which is equivalent to $\bar{n}P_g = \mathrm{const}$ given the assumption~\eqref{eq:bias-evolution} for the redshift evolution of the linear bias, or that we obtain spectra evenly across the sky and along the radial direction.\medskip

\begin{figure}
	\centering
	\includegraphics{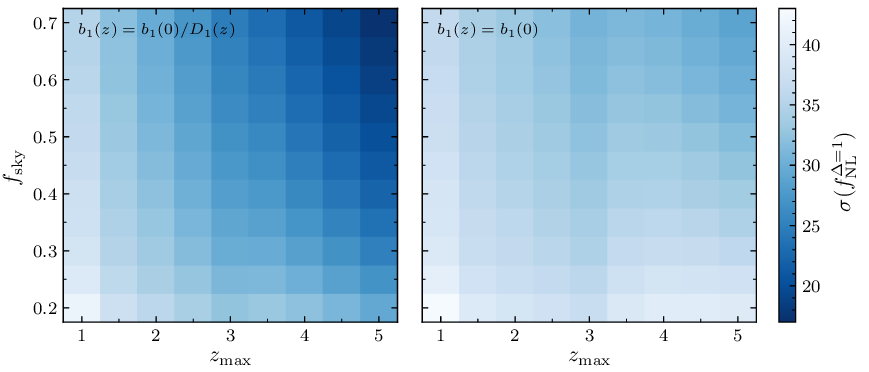}\vspace{-5pt}
	\caption{Dependence of $\sigma(\fnl^{\Delta=1})$ on the sky fraction~$\fsky$ and the maximum redshift~$\zmax$ at fixed observational effort for the double-tracer billion-object survey. The left panel shows the fiducial case, $b_1(z) \propto D(z)^{-1}$~[cf.~\eqref{eq:bias-evolution}] and the right panel displays the constant bias case, $b_1(z) = \mathrm{const}$. While we only show the results for $\Delta = 1$, the displayed trends are similar for all scaling exponents~$\Delta$.}
	\label{fig:zmax}
\end{figure}

Higher-redshift objects are generally expected to yield larger values of~$b_1(z)$~[see the discussion around~\eqref{eq:bias-evolution}]. Since $b_\phi(z) \propto (b_1(z)-1)$ from the universality relation~\eqref{eq:universality}, the signal can be enhanced by a large amount by extending the survey to redshifts where $b_1(z) > 2$ instead of $b_1(z) \approx 1$. Assuming the bias evolution model of~\eqref{eq:bias-evolution}, we see in Fig.~\ref{fig:zmax}%
that we maximize our sensitivity to~$\fnl^\Delta$ even at constant effort by maximizing~$\zmax$. While we only show the results for $\Delta = 1$, the qualitative features in the figure are independent of the scaling exponent~$\Delta$. This implies that increasing the redshift range at fixed observational effort increases the constraining power for the entire range of~$\Delta$.

Increasing the redshift range also increases the number of modes. If we however hold $b_1(z) = b_1$ fixed and increase~$\zmax$, we see in the right panel of Fig.~\ref{fig:zmax} that there is a more complex relationship between~$\zmax$ and the sensitivity to~$\fnl^\Delta$. For instance, keeping $\zmax = 3$ and increasing~$\fsky$ also increases the number of modes without the added observing time per object needed at high redshifts. Having said that, the benefit of a large sky fraction is much weaker than the impact of larger~$b_1$ at high~$z$ which can be seen by the lower overall sensitivity to~$\fnl^\Delta$ in the right panel.

By comparing the two panels in Fig.~\ref{fig:zmax}, we can see that the optimal configuration is still at the largest maximum redshift and sky fraction, $\zmax = 5$ and $\fsky = 0.7$, for either biasing model. For smaller~$\fsky$, the large advantage offered by larger~$\zmax$ however disappears when going from evolving to fixed bias due to the lower overall number density of galaxies at higher redshifts. In this case, more emphasis is placed on high-redshift targets without the corresponding boost in the signal. (The drop-off in~$\sigma(\fnl)$ between $\zmax = 3.0\text{ and }3.5$ is the result of our simplified choice of dividing the observational effort at $z = 3$.)\medskip

To summarize, at fixed observational effort, increasing the redshift range via a larger~$\zmax$ has the largest influence on the sensitivity of a survey. The reason is that~$\zmax$ increases both the size of the signal and the number of modes. The former is particularly important since it offsets the increased noise associated with a large-volume survey. We however note that this conclusion does depend on the redshift dependence of the bias and the observing time needed for acquiring spectra of these high-redshift objects.

\subsubsection*{Photometric versus Spectroscopic Surveys}

The benefit of a three-dimensional survey can simply be estimated by counting modes for many cosmological parameters, including~$\fnl$ from the bispectrum. The number of modes in a three-dimensional survey scales as~$\kmax^3$, while it scales as~$\kmax^2$ for a two-dimensional survey. The benefits of a three-dimensional survey are however less clear for scale-dependent bias since the information for this signal manifests itself at low~$k$, in particular for $\Delta \lesssim 0.5$. In reality, the trade-off is of course not between a two- and a three-dimensional survey, but between surveys which measure redshifts spectroscopically and photometrically. While photometric redshifts are less precise, they may be good enough to provide three-dimensional information on the scales needed for scale-dependent bias for a significant range of~$\Delta$ which we investigate now.\footnote{Spectroscopic surveys will have additional benefits for controlling systematics, such as projecting out large-scale effects, but typically at the cost of larger shot noise. Quantifying the importance of spectra in this context is a delicate issue which is beyond the scope of this work.}\medskip

\begin{figure}
	\centering
	\includegraphics{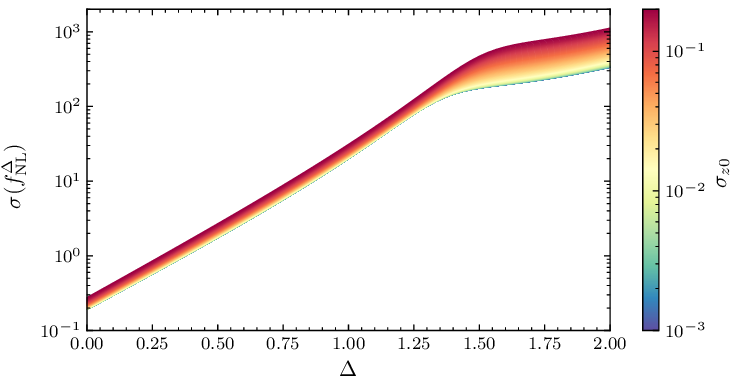}\vspace{-7pt}
	\caption{Comparison of~$\sigma(\fnl^\Delta)$ for a range of photometric redshift errors~$\sigma_{z0} \in [0.001, 0.2]$ for the billion-object survey. We observe that photometric redshift errors only mildly degrade the constraining power of a survey for $\Delta \lesssim 1.3$. Photometric surveys therefore seem to be able to access most of the information about the scale-dependent bias with the additional benefit of generally lower noise levels in the galaxy power spectrum.}
	\label{fig:photoz}
\end{figure}

The impact of redshift errors on the constraining power of the billion-object survey is shown in~Fig.~\ref{fig:photoz}.%
We see, somewhat surprisingly, that spectroscopic surveys do not provide a significant improvement to the power spectrum measurement of~$\fnl^\Delta$ for $\Delta \lesssim 1-1.3$. (The specific value of this transition in~$\Delta$ depends on the employed biasing model.) In practice, photometric redshifts therefore appear to be good enough to access a lot of the available information about the scale-dependent bias. This conclusion is supported by the similarity of the forecasts for~SPHEREx and~LSST. While the two surveys pursue very different strategies for obtaining redshifts, the resulting forecasts are quite similar. At first sight, this similarity is still surprising because it does not reflect the impact of the difference in number density shown in Fig.~\ref{fig:nbar}. The reason for this, however, is just that multi-tracer analyses are not implemented in conventional LSST~forecasts. This naturally raises the question of whether a multi-tracer analysis would lead to an advantage of~LSST over some spectroscopic surveys given its very high number density~(cf.\ Fig.~\ref{fig:future_surveys}). We will explore this in detail in Section~\ref{sec:multitracer}.\medskip

Importantly, we have not discussed the impact of systematic effects~\cite{Pullen:2012rd} on photometric and spectroscopic surveys. Given that the signals are the largest on large angular scales~(especially for $\Delta \lesssim 0.5$), a variety of atmospheric, detector and astrophysical effects can contaminate our signal. Having said that, we expect that many of the conclusions drawn from these forecasts will be robust over a large parameter range since the scales on which the signals dominate vary with the scaling exponent~$\Delta$.

\section{Constraints and Data Analysis for General Scaling Exponents}
\label{sec:analysis}

In the previous section, we focused on forecasting current and future constraints on~$\fnl^\Delta$, and on how to optimize a survey to measure~PNG from the galaxy power spectrum for $\Delta \in [0, 2]$. In this section, we will apply this knowledge to constrain~$\fnl^\Delta$ over this same range of scaling exponents~$\Delta$ using the BOSS~DR12~galaxy power spectra. We will first derive and propose an effective and convenient search strategy. We will then perform this measurement and compare it to a direct analysis.

\subsection{Search Strategy}
\label{sec:strategy}

The scaling exponent~$\Delta$ is fixed in some inflationary models, while it is a free parameter in others. For the standard local and equilateral shapes, the value of~$\Delta$ are fixed by their templates to $\Delta = 0$ and $\Delta = 2$, respectively. The physics of these cases is clear since they arise from multi- and single-field dynamics. In quasi-single-field inflation, we have $\Delta = 3/2 - \sqrt{9/4 - m^2/H^2}$ which implies that it is natural to scan over values of $\Delta \in [0, 3/2]$ in an analysis of those models. The natural question we face in a practical analysis of current data therefore is whether it is prudent to vary~$\fnl^\Delta$ and $\Delta \in [0, 2]$ as free parameters, or whether we should hold the scaling exponent fixed to some value as suggested by a given model, $\Delta = \Delta'$, and only vary~$\fnl^{\Delta'}$.\medskip

The strategy we will follow is to hold~$\Delta$ fixed to a few discrete values, e.g.~$\Delta = 0, 2$ or $\Delta = 0.0, 0.1, \ldots, 1.9, 2.0$, and infer the associated constraints on~$\fnl^\Delta$ at each respective~$\Delta$. The reason why we prefer this strategy over scanning over both parameters simultaneously is that small changes to~$\Delta$ have no impact on the signal to noise and, therefore, represent a degenerate direction in the analysis. This becomes evident in the Fisher matrix formalism, where we have for a fiducial value of $\fnl^\Delta = 0$:
\begin{equation}
	\left.\frac{\partial}{\partial \Delta} P_{gg}(k)\right|_{\fnl^\Delta = 0} = 6 \fnl^\Delta b_\phi(k)\, k^{-2} \T(k)^{-1} \log(k R_\star) (k R_\star)^\Delta\, b(k) P_\mathrm{lin}(k) \Big|_{\fnl^\Delta = 0} = 0 \, .
\end{equation}
We therefore have no reason to expect marginalizing over~$\Delta$ would lead to more accurate results until our data are comfortably excluding $\fnl^\Delta = 0$.

We can quantify to what degree the data can distinguish two values of the scaling exponent, $\Delta_1$ and $\Delta_2$, through the ``cosine'' between the signals,
\begin{equation}
	\cos{(\fnl^{\Delta_1},\fnl^{\Delta_2})} = \frac{F_{\fnl^{\Delta_1} \fnl^{\Delta_2}}}{\sqrt{F_{\fnl^{\Delta_1} \fnl^{\Delta_1}}F_{\fnl^{\Delta_2} \fnl^{\Delta_2}}}} \equiv \frac{F_{12}}{\sqrt{F_{11} F_{22}}} \, ,	\label{eq:cos-f1-f2}
\end{equation}
where we defined $F_{ij} \equiv F_{\fnl^{\Delta_i},\fnl^{\Delta_j}}$. This is precisely the same definition of the cosine used to define the PNG~shapes in a bispectrum analysis~\cite{Babich:2004gb}. We can also generalize this to the Fisher matrix after marginalizing over the bias parameters by inverting the Fisher matrix, truncating to $\fnl^{\Delta_1}$ and $\fnl^{\Delta_2}$, and inverting the truncated matrix to get the effective Fisher matrix for these two parameters. We therefore take~$F_{ij}$ to be the marginalized, two-dimensional Fisher matrix for the parameters $\fnl^{\Delta_1}$ and $\fnl^{\Delta_2}$ from now on.

\begin{figure}
	\centering
	\includegraphics{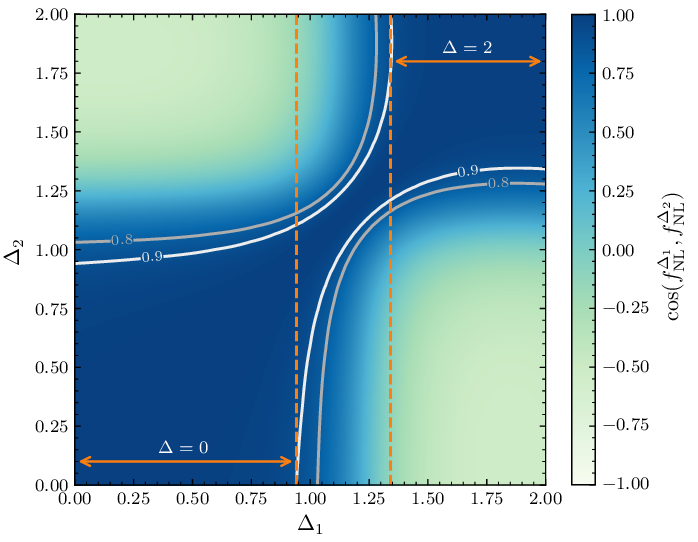}\vspace{-7pt}
	\caption{Correlation matrix for measurements of galaxy power spectra with different values of the non-Gaussian exponent~$\Delta$ as defined in~\eqref{eq:cos-f1-f2}. This was computed for~BOSS with fiducial values of $\fnl^{\Delta_i} = 0$, fixed $\Lambda$CDM~parameters and marginalized over all biases. The orange arrows indicate the coverage of the local~($\Delta = 0$) and equilateral~($\Delta = 2$) templates with a correlation coefficient of larger than~$0.9$.}
	\label{fig:cos-f1-f2}
\end{figure}

The correlation~\eqref{eq:cos-f1-f2} between the measurements of~$\fnl^{\Delta_1}$ and~$\fnl^{\Delta_2}$ is shown in Fig.~\ref{fig:cos-f1-f2}%
using the experimental specifications for the BOSS~survey of Table~\ref{tab:boss-specs} which have yielded forecasted results consistent with performed data analyses in the past~(see e.g.~\cite{Beutler:2019ojk}).\footnote{Note that for the calculation of the correlation matrix for~BOSS, we assumed fiducial values of~$b_{s^2}$ and~$b_{\Pi\Pi^{[2]}}$ based on the Lagrangian local-in-matter-density biasing model~\cite{Desjacques:2016bnm} and $b_{\delta^2}$~based on halo simulation fit in~\cite{Lazeyras:2015lgp}. This choice~(rather than setting them to~$0$ as in most other cases in this paper) leads to minimal differences in the correlation coefficient. In addition, we impose the same Gaussian priors on the loop biases as quoted in~\cite{Philcox:2021kcw} which barely affect the forecasted constraints for small~$\Delta$ and up to a factor of four at large~$\Delta$, as expected. Finally, we employ a slightly larger minimum wavenumber $\kmin = \SI{0.01}{\hPerMpc}$. All these choices are guided by the data analysis performed in~\textsection\ref{sec:BOSS}.} The cosine computed for other surveys is similar~(see Appendix~\ref{app:scaling}). The key take-away from this correlation is that we can effectively constrain almost the entire range $\Delta \in [0, 2]$ using measurements only at $\Delta = 0\text{ and }2$. Specifically, using the Cauchy-Schwartz inequality and the Cram\'er-Rao~(CR) bound, we have
\begin{equation}
	(F_{22})^{-1} \leq \frac{F_{11}}{F_{12}^2} \leq \frac{\sigma^2(\fnl^{\Delta_1})}{\cos^2(\fnl^{\Delta_1},\fnl^{\Delta_2})} \frac{F_{11}}{F_{22}} \, ,
\end{equation}
where~$\sigma^2(\fnl^{\Delta_1})$ is the variance of the measurement of~$\fnl^{\Delta_1}$ while holding $\fnl^{\Delta_2}$~fixed. The best possible constraint we could place on~$\fnl^{\Delta_2}$, $\sigma^2(\fnl^{\Delta_2})_{\mathrm{CR}} = F_{22}^{-1}$, is therefore bounded by the measurement of~$\fnl^{\Delta_1}$ via
\begin{equation}
	\sigma^2(\fnl^{\Delta_2})_\mathrm{CR} \leq \frac{\sigma^2(\fnl^{\Delta_1})}{\cos^2(\fnl^{\Delta_1},\fnl^{\Delta_2})} \frac{F_{11}}{F_{22}} \, .	\label{eq:bound_derived}
\end{equation}
As a result, we can place an approximate constraint on~$\fnl^{\Delta_2}$ by rescaling the measurement of~$\fnl^{\Delta_1}$ using the elements of the marginalized Fisher matrix~$F_{ij}$.\medskip

It is important to highlight that this strategy is only effective if our parameter inference is consistent with $\fnl^\Delta = 0$. To be more precise, all values of~$\Delta$ look similar at low signal to noise for~$\fnl^\Delta$. On the other hand, the data is no longer independent of~$\Delta$ if one or more of your~$\fnl^\Delta$ measurements at specific~$\Delta$ show significant evidence for $\fnl^\Delta \neq 0$. In that case, a measurement at a different value~$\Delta'$ could produce a much larger or much smaller signal to noise. We refer to Appendix~\ref{app:scaling} for more details on this scenario.

\subsection{BOSS DR12 Analysis}
\label{sec:BOSS}

Based on the strategy laid out above, we inferred constraints on~$\fnl^\Delta$ from the BOSS~DR12 dataset following the general setup of the analysis for local, orthogonal and equilateral~PNG performed in~\cite{Cabass:2022wjy, Cabass:2022ymb}~(see also~\cite{DAmico:2022gki}) with a few different~(generally more conservative) choices. In the following, we provide a brief overview of our data analysis, describe these differences and report the limits on~PNG that we inferred directly and via the correlation matrix~\eqref{eq:cos-f1-f2}.\medskip

We analyzed the power spectrum from galaxy clustering data of the twelfth and final release of~BOSS, referred to as~DR12~\cite{BOSS:2016wmc}. This dataset contains the positions of about \num{1.2e6}~galaxies between redshifts~$0.2$ and~$0.75$ in a cosmic volume of roughly $\SI{5.8}{\per\h\cubed\Gpc\cubed}$ divided into four subsets: the Northern and Southern galactic caps of the two non-overlapping redshift bins~$[0.2, 0.5]$ and~$[0.5, 0.75]$ with effective mean redshifts $z = 0.38\text{ and }0.61$. We employ the galaxy power spectrum multipoles~$P_{gg}^{(\ell)}\hskip-1pt(k)$, with $\ell = 0, 2, 4$, measured using a quadratic window-function-free estimator~\cite{Philcox:2020vbm}. The corresponding covariance matrices were computed from MultiDark-Patchy mock catalogs~\cite{Kitaura:2015uqa}. This follows the analyses in~\cite{Cabass:2022wjy, Cabass:2022ymb} restricted to the power spectrum multipoles. We model the nonlinear galaxy power spectrum in redshift space as implemented in~\texttt{CLASS-PT}~\cite{Chudaykin:2020aoj}~(based on~\texttt{CLASS}~\cite{Blas:2011rf}) which follows the analyses of~\cite{Ivanov:2019pdj, DAmico:2019fhj} based on the effective field theory of large-scale structure~(EFTofLSS)~\cite{Baumann:2010tm, Carrasco:2012cv, Pajer:2013jj, Carrasco:2013mua} (see~\cite{Cabass:2022avo, Ivanov:2022mrd} for recent reviews) and galaxy bias expansion~\cite{Desjacques:2016bnm}, including the Alcock-Paczynski effect. We additionally include the non-Gaussian contributions from the scale-dependent bias~\eqref{eq:bias_delta} due to $\fnl^\Delta \neq 0$ similar to the analyses in~\cite{Cabass:2022wjy, Cabass:2022ymb}.\footnote{In this paper, we have not considered the imprints of light inflationary fields in the bispectrum, but focused on the scale-dependent bias in the power spectrum. To remain model agnostic, we also did not include the contribution of a non-zero primordial bispectrum to the galaxy power spectrum.}

We performed a Markov~chain Monte~Carlo~(MCMC) analysis of the data using the Metropolis-Hastings sampler of \texttt{MontePython}~\cite{Audren:2012wb, Brinckmann:2018cvx}, varying the theoretical galaxy power spectrum multipoles in each step. We fix the $\Lambda$CDM~parameters to the Planck~2018 best-fit values~\cite{Planck:2018vyg}~(with the sum of neutrino masses $\sum m_\nu = \SI{0.06}{eV}$) instead of including Planck information on~$\Lambda$CDM since we saw essentially no change in the forecasted constraints on~$\fnl^\Delta$ of Section~\ref{sec:forecasts} when marginalizing over all of our galaxy bias parameters. Per redshift bin and galactic cap, we however separately vary the bias parameters~$b_1^{(i)}$, $b_2^{(i)}$ and~$b_{\mathcal{G}_2}^{(i)}$ as defined in~\cite{Assassi:2014fva, Senatore:2014eva},\footnote{The bias expansions used in our forecasts and data analysis are equivalent.} and analytically marginalize over the other eight biases and EFTofLSS~counterterms which appear linearly in the theoretical galaxy power spectrum~\cite{Philcox:2020zyp}. We impose flat priors on the linear biases~$b_1^{(i)} \in [1, 4]$ and the non-Gaussian amplitude~$\fnl^\Delta$~(infinitely wide), and Gaussian priors on the other nuisance parameters as described in~\cite{Philcox:2021kcw}. Following~\cite{Cabass:2022wjy}, we also marginalize over the non-Gaussian bias $b_\phi \to N_{b_\phi} b_\phi$ separately for each data subset and impose a wide Gaussian prior on its normalization, $N_{b_\phi} \sim \mathcal{N}(1, 5)$, which is motivated by the peak-background split model~\cite{Schmidt:2010gw}.\footnote{It might appear surprising that we marginalize over~$b_\phi$ with a Gaussian prior given the degeneracy with~$\fnl^\Delta$. Note that the mean value of the Gaussian is still set by the universality relation~\eqref{eq:universality} with $p=1$, but its variance is very wide. In addition, we marginalize over~$b_\phi$ in each of the four data subsets separately, as we do with the Gaussian bias parameters. This procedure alleviates degeneracies between~$\fnl^\Delta$ and~$b_1$ which can be severe if we assume that the universality relation is exact, especially for $\Delta \sim 1$. This procedure is an imperfect solution to address the uncertainty in~$b_\phi$ which we plan to address as part of~\cite{Green:inprep}.} On the other hand, the non-Gaussian parameter~$\fnl^\Delta$ is commonly varied for all four data subsets at fixed exponent~$\Delta$. We (more)~conservatively limit the range of wavenumbers to $k \in \SIrange{0.01}{0.13}{\hPerMpc}\text{ and } \SIrange{0.01}{0.16}{\hPerMpc}$ for the low- and high-redshift bin, respectively, after initial tests. We therefore employ a smaller maximum wavenumber than in~\cite{Cabass:2022ymb, Cabass:2022wjy}, who tested the validity of the analysis for $\Delta = 0\text{ and }2$ on mock catalogs, and a slightly higher~(same) minimum~(maximum) wavenumber than the forecasts shown in Fig.~\ref{fig:future_surveys}. Since $\Delta k = \SI{0.005}{\hPerMpc}$, this choice implies that we employ 24~and 30~$k$-bins for the respective redshift bins, each power spectrum multipole and both galactic caps. All chains converged with a Gelman-Rubin criterion of $R-1 < 0.01$~(usually much smaller) for each parameter.\medskip

\begin{figure}
	\centering
	\includegraphics{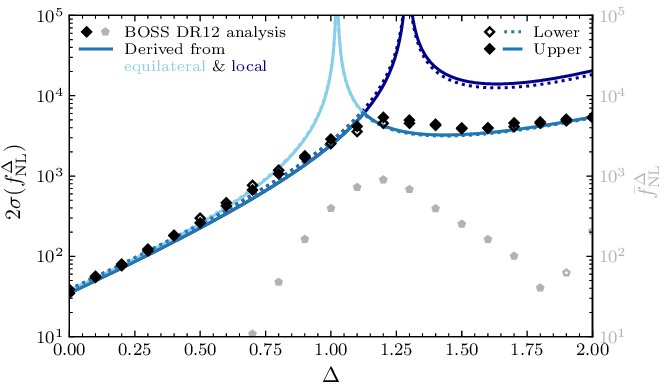}\vspace{-7pt}
	\caption{Comparison of the measured~(diamonds) and derived~(lines) values for twice the standard deviation of~$\fnl^\Delta$, $2\sigma(\fnl^\Delta)$, for the BOSS~DR12 dataset. The light~(dark) blue lines show the approximate value of~$2\sigma(\fnl^\Delta)$ derived from the data with $\Delta = 2~(0)$, i.e.\ the equilateral (local)~PNG templates using~\eqref{eq:constraint_derived}, with the minimum of these being shown in blue. The diamonds indicate the constraints directly inferred from BOSS~DR12 for fixed values of~$\Delta$, while the gray pentagons are the corresponding mean values. Filled~(unfilled) symbols and solid~(dotted) lines represent positive or upper (negative or lower) values. We do not find evidence for $\fnl^\Delta \neq 0$ for any $\Delta \in [0, 2]$, and generally good agreement between the measured and derived limits on general~PNG. This demonstrates that our proposed search strategy works on data.}
	\label{fig:BOSS}
\end{figure}

The results of our separate MCMC~data analyses for 21~fixed values of the non-Gaussian exponent~$\Delta$ are shown in Fig.~\ref{fig:BOSS}%
and we provided the numerical constraints for $\Delta=0.0, 0.5, 1.0, 1.5, 2.0$ in~\eqref{eq:boss-constraints}. Since the posterior distributions for~$\fnl^\Delta$ are slightly non-Gaussian, we display the mean values in gray pentagons, and the upper and lower $2\sigma$~constraints in filled and unfilled black diamonds, respectively. We do not find any evidence for any PNG~shape with $\Delta \in [0, 2]$ and observe the same characteristic functional dependence of~$2\sigma(\fnl^\Delta)$ on~$\Delta$ as we forecasted in Fig.~\ref{fig:future_surveys}.\footnote{The non-Gaussian amplitude~$\fnl^\Delta$ is defined relative to the scalar amplitude, $A_\mathrm{s}^{1/2} \approx \num{5e-5}$~\cite{Planck:2018vyg}, such that weak~PNG corresponds to $\fnl A_\mathrm{s}^{1/2} \ll 1$. This means that our bounds $2\sigma(\fnl^\Delta) \sim \num{e4}$ are approaching the limits of the range of validity of our implicit expansion in linear order in~$\fnl^\Delta$. This concern is however alleviated even with a very conservative inclusion of bispectrum data~\cite{Cabass:2022wjy, DAmico:2022gki}.} We note that the difference in forecasted constraining power of Fig.~\ref{fig:future_surveys} and the data analysis of Fig.~\ref{fig:BOSS} can be mainly attributed to the use of Gaussian priors on the higher-order galaxy bias parameters in Section~\ref{sec:analysis}. In fact, we find reasonable consistency if we include the priors on the Gaussian biases in our Fisher matrices as discussed in~\textsection\ref{sec:strategy}. Compared to~\cite{Cabass:2022wjy, Cabass:2022ymb, DAmico:2022gki}, our constraints are weaker, as expected due to our more conservative approach and use of less data, but consistent. This is particularly evident for the case of the equilateral shape~($\Delta = 2$) that dramatically benefits from bispectrum information in various ways which is absent in our analysis of the power spectrum alone. On the other hand, our limits on~$\fnl^\Delta$ are comparable to the constraints on~$\fnleq$ when including bispectrum information~\cite{Cabass:2022wjy} for $\Delta \lesssim 0.75$. While the bispectrum will always add information, we might however \mbox{imagine there is a more limited gain in this regime.}

We can also compare our bounds on~$\fnl^\Delta$ to those obtained in previous analyses of CMB~and LSS~data. As already anticipated and explained in Section~\ref{sec:forecasts}, the CMB~limits on local and equilateral~PNG from the Planck~2018 bispectra are significantly better~\cite{Planck:2019kim}. This is also the case for the constraints for $\Delta \in [0, 3/2]$ derived from Planck~2013 data~\cite{Planck:2013wtn}, $|\fnl^\Delta| \lesssim 50$, which is dominated by the equilateral bispectrum while our bounds arise from the squeezed limit of the quasi-single-field shapes. The only previous analysis constraining~$\fnl^\Delta$ from LSS~data was performed in~\cite{Agarwal:2013qta} on SDSS-III~DR8 data with more limited modeling and range of scales. It is therefore unsurprising that our bounds are considerably more constraining.\medskip

In addition to directly inferring constraints on~$\fnl^\Delta$ from the data, there exists a second path to inferring the bounds at a given value of~$\Delta$ based on the correlation of different values of~$\Delta$ in the marginalized galaxy power spectrum as discussed in~\textsection\ref{sec:strategy}. We take $\Delta = \Delta_1 = 0\text{ and }2$ as the fiducial values for this inference since these scaling exponents correspond to the signal from local and equilateral~PNG, respectively, which are usually considered in data analyses. Based on~\eqref{eq:bound_derived}, we obtain these derived bounds as follows:
\begin{equation}
	\left[2\sigma(\fnl^\Delta)\right]_\mathrm{MCMC} = \left[\frac{1}{\cos(\fnl^\Delta, \fnl^{\Delta=0,2})} \frac{\sigma(\fnl^\Delta)}{\sigma(\fnl^{\Delta=0,2})}\right]_\mathrm{Fisher} \left[2\sigma(\fnl^{\Delta=0,2})\right]_\mathrm{MCMC}\, ,	\label{eq:constraint_derived}
\end{equation}
where the subscript `Fisher'~(`MCMC') indicates that these quantities are calculated based on the Fisher matrix~(from the MCMC~analysis). The correlation term~$\cos(\fnl^\Delta, \fnl^{\Delta=0,2})$ is displayed in~Fig.~\ref{fig:cos-f1-f2}, while the standard deviations~$\sigma(\fnl^\Delta)$ are computed from the marginalized one-dimensional Fisher matrix as in Section~\ref{sec:forecasts}, but including the same Gaussian priors as in the correlation term and the data analysis. The constraints derived from the measurement at $\Delta=0\text{ and }2$ are displayed in Fig.~\ref{fig:BOSS} as the dark and light blue curves. We see that the constraint derived from~$2\sigma(\fnl^{\Delta=0})$ dominates for $\Delta \lesssim 1.1$, while the constraint derived from $2\sigma(\fnl^{\Delta=2})$ dominates for larger~$\Delta$, which is consistent with our expectations from Fig.~\ref{fig:cos-f1-f2}.

Overall, we find generally reasonable agreement between the directly inferred limits on~$\fnl^\Delta$, and those jointly derived from the local and equilateral measurements. This is especially true for those values of~$\Delta$ with large correlation and negligible mean values. This implies that we can use~\eqref{eq:constraint_derived}, with the correlation term calculated in a Fisher matrix approach for a given survey, to derive~$2\sigma(\fnl^{\Delta\neq0,2})$ in the absence of a detection of $\fnlloc,\fnleq \neq 0$. In other words, we have demonstrated that the search strategy that we proposed in~\textsection\ref{sec:strategy} works on BOSS~data and allows to dramatically reduce the number of analyses needed to constrain general forms of primordial non-Gaussianity and light inflationary fields as parametrized by~$\Delta$ from measurements of local and equilateral non-Gaussianity. The agreement between our analysis and forecasts, together with the consistency with prior analyses, is an additional sign that searches for light fields with the galaxy power spectrum will be competitive with the~CMB for near-term galaxy surveys.

\section{Multiple Tracers and the Dependence on Astrophysics}
\label{sec:multitracer}

Sample-variance cancellation plays an important part in improving the measurement of primordial non-Gaussianity from the galaxy power spectrum. Applying this technique however requires knowledge of the details of the specific galaxy samples used in the multi-tracer forecast or analysis. In this section, we will first investigate how an optimal survey design can maximize the scientific return of multiple tracers in spectroscopic surveys. Then, we will consider the untapped potential of a multi-tracer configuration of~LSST and compare it to~SPHEREx.

\subsection{Optimizing Spectroscopic Multi-Tracer Analyses}

We now explore the potential of multi-tracer analyses and how astrophysical details of the galaxy samples affect the constraints on~$\fnl^\Delta$. Given our biasing model~\eqref{eq:galaxy-overdensity} with~\eqref{eq:scale-dependent_bias}, this comes down to the description of the non-Gaussian bias parameter~$b_\phi(z)$, which depends on three parameters when using the universality relation~\eqref{eq:universality}: $\delta_c$, $b_1(z)$ and~$p$. Since the parameter~$\delta_c$ is completely degenerate with~$\fnl^\Delta$, we will focus on exploring what combination of~$b_1(z)$ and~$p$ leads to the biggest improvement on~$\sigma(\fnl^\Delta)$.\medskip

For galaxy power spectrum forecasts to recover all the information in the primordial statistics, we need to apply the multi-tracer technique and its ability for sample-variance cancellation at very high number density. For example, we know the information in the primordial bispectrum for $\Delta > 0$ is dominated by $k \sim \kmax$~\cite{Babich:2004gb}. On the other hand, the non-Gaussian information in the single-tracer power-spectrum forecasts are limited by sample variance at small wavenumbers for $\Delta \lesssim 0.5$. Sample-variance cancellation eliminates this artificial dependence on the noise at low~$k$, limiting a measurement instead by the shot noise associated with galaxy formation~\cite{dePutter:2018jqk}, which corresponds to modes at $\kmax \sim k_\mathrm{halo}$.

Realistic spectroscopic surveys are unfortunately far from being in the fully multi-tracer regime as we have seen in Fig.~\ref{fig:nbar}. As a result, it is harder to get reliable intuition for how to optimize the measurement of~$\fnl^\Delta$. We have previously explored how to maximize the science of a spectroscopic survey by changing the experimental configuration in terms of~$\fsky$, $\zmax$, etc. Here, we will assume a fixed survey geometry and investigate what additional information can be extracted by taking advantage of sample-variance cancellation.\medskip

\begin{figure}
	\centering
	\includegraphics{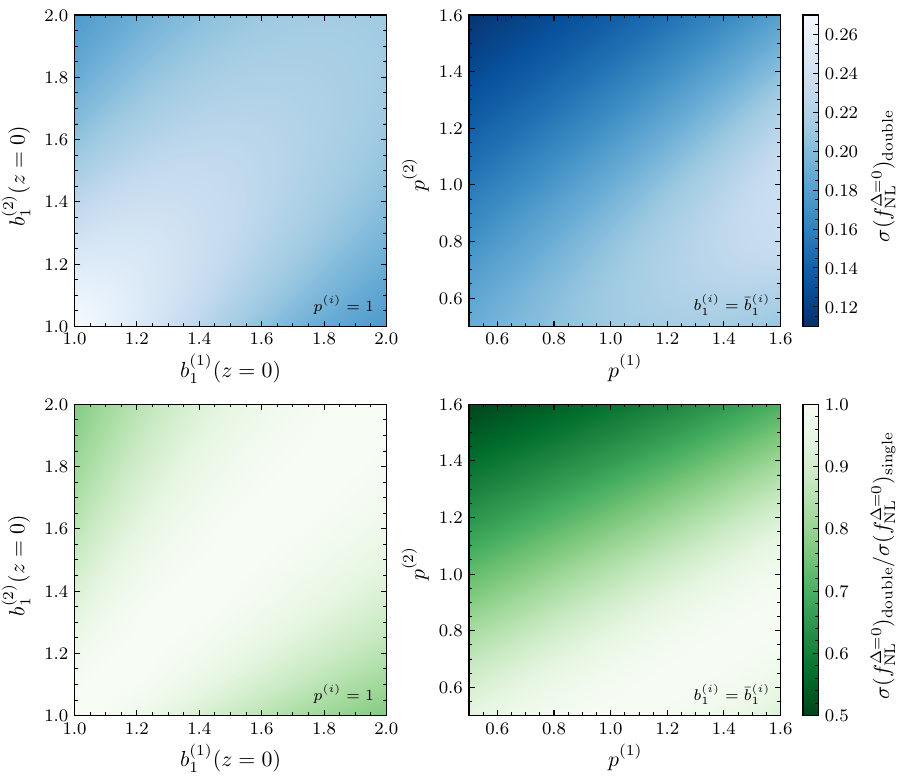}\vspace{-7pt}
	\caption{Double-tracer forecast for $\sigma(\fnl^\Delta)$ with $\Delta = 0$ for the billion-object survey~(\textit{top}) and relative comparison to a single-tracer forecast, $\sigma(\fnl^\Delta)_\mathrm{double}/\sigma(\fnl^\Delta)_\mathrm{single}$~(\textit{bottom}), with fixed $\Lambda$CDM~parameters. In the left column, we vary the linear biases~$b_1^{(i)}$ of the two samples at fixed $p^{(i)}_{\vphantom{1}} = 1$, i.e.\ similar to our forecasts in Section~\ref{sec:forecasts}. In the right column, we use the fiducial biases of the samples~[cf.~\eqref{eq:bias-evolution}], with $\bar{b}_1^{(1)}(z=0) = 2.0 > 1.2 = \bar{b}_1^{(2)}(z=0)$, and vary the parameter~$p^{(i)}_{\vphantom{1}}$ in the universality relation for~$b_\phi^{(i)}$~[cf.~\eqref{eq:universality}]. While we only show these results for $\Delta = 0$, the qualitative features of these panels remain the same for other values of the scaling exponent.}
	\label{fig:multi-double}
\end{figure}

In the fully multi-tracer regime, the inverse variance depends on the linear bias~$b_1^{(i)}$ and non-Gaussian bias~$b_\phi^{(i)}$ as~\cite{Barreira:2023rxn}
\begin{equation}
	\sigma(\fnl^\Delta)^{-2} \propto \left|b_1^{(1)} b_\phi^{(2)} - b_1^{(2)} b_\phi^{(1)}\right| .	\label{eq:multitracer-bias-scaling}
\end{equation}
On the other hand, we simply maximize~$b_1 b_\phi$ in the single-tracer regime. The question therefore is how we can choose the samples to minimize~$\sigma(\fnl^\Delta)$ in realistic surveys. In Figure~\ref{fig:multi-double},\footnote{Figure~\ref{fig:multi-double} only shows the forecasted constraints and improvements for the scaling exponent $\Delta = 0$, i.e.\ local~PNG. When we marginalize over all biases, the scaling behavior of the scale-dependent bias however has little effect on the improvement factor and the marginalized biasing model only changes the overall constraining power as indicated by the color bar in the upper panels. In other words, the overall qualitative behavior conveyed in Fig.~\ref{fig:multi-double} remains intact for different biasing models and scaling exponents.}%
we show the relative comparison of~$\sigma(\fnl^\Delta)$ between a double- and single-tracer forecast of the billion-object survey while holding the $\Lambda$CDM parameters fixed.\footnote{When marginalizing over the biases, the difference between the forecasts for fixed and marginalized $\Lambda$CDM~parameters~(with a Planck prior) is at most a few percent. We therefore fix the $\Lambda$CDM~parameters to their fiducial values for all forecasts in this section.} We show this for $\Delta = 0$~(see also e.g.~\cite{Karagiannis:2023lsj}) since we expect this case to benefit the most from a multi-tracer analysis, but explicitly verified that the forecasts with $\Delta > 0$ show at most comparable levels of improvement.

We vary the linear biases of the two subsamples of the survey in the left column of Fig.~\ref{fig:multi-double}, while fixing $p = 1$, as in all of our forecasts in Section~\ref{sec:forecasts}. Note that we only vary the bias at $z = 0$, as labeled in the figure, with the biases of the actual galaxy samples then being calculated based on the bias evolution model of~\eqref{eq:bias-evolution}. The values along the diagonal $b_1^{(1)} = b_1^{(2)}$ correspond to a single-tracer forecast, in which~$\sigma(\fnl^\Delta)$ simply decreases with increasing~$b_1^{(i)}$. We observe that the multi-tracer improvement is most pronounced here when the difference between~$b_1^{(1)}$ and~$b_1^{(2)}$ is maximized. We additionally see that the improvement of the double- over the single-tracer approach is limited to about~30\%.\medskip

So far, we have assumed a universal halo mass function which corresponds to $p = 1$ in the universality relation~\eqref{eq:universality}. Even though this has been widely adopted in the literature, this assumption is not a perfect description of halos in simulations and will likely not be generally true for actual galaxy samples, cf.\ e.g.~\cite{Desjacques:2008vf, Pillepich:2008ka, Grossi:2009an, Reid:2010vc, Wagner:2011wx, Hamaus:2011dq, Scoccimarro:2011pz, Baldauf:2015vio, Biagetti:2016ywx, Barreira:2020kvh}.\footnote{We note that, as long as $b_\phi \neq 0$, a detection of scale-dependent bias for a single redshift bin and tracer implies a detection of primordial non-Gaussianity, $\fnl^\Delta \neq 0$. However, reaching the full sensitivity of a survey requires combining multiple tracers and/or redshift bins which necessitates knowledge of how~$b_\phi(z)$ changes across the samples. Most analyses will therefore make assumptions about~$b_\phi$, such as the universality relation~\eqref{eq:universality} and the values of~$p$.} This is why we now explore the possibility of improving the constraints on~$\fnl^\Delta$ by selecting tracers within the parametrization of~$b_\phi$ through~$p$~(assuming such knowledge is already reliably known from simulation results).

\begin{figure}
	\centering
	\includegraphics{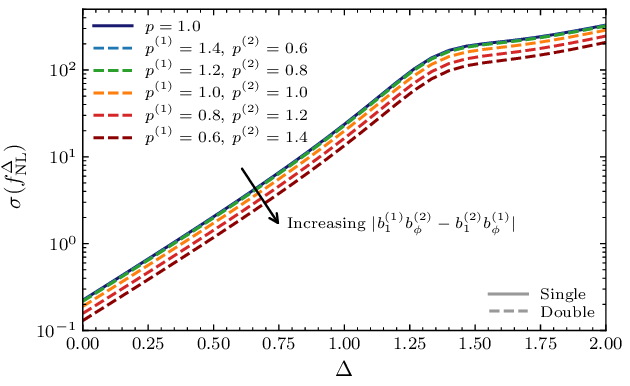}\vspace{-7pt}
	\caption{Forecasted constraints on~$\fnl^\Delta$ for different values of the parameter~$p^{(i)}$ in the universality relation~\eqref{eq:universality} of~$b_\phi$ for the double-tracer configuration of the billion-object survey and fixed $\Lambda$CDM~parameters. We observe that~$\sigma(\fnl^\Delta)$ improves for all~$\Delta$ when $|b_1^{(1)} b_\phi^{(2)} - b_1^{(2)} b_\phi^{(1)}|$ increases which is the direction indicated by the arrow.}
	\label{fig:multi-p}
\end{figure}

We therefore vary the parameters~$p^{(i)}$ of the two galaxy samples while using their fiducial biases in the right column of Fig.~\ref{fig:multi-double}. Note that the fact that the linear biases for the two tracers are different, with $b_1^{(1)}(z = 0) = 2.0$ and $b_1^{(2)}(z = 0) = 1.2$, is the reason for the asymmetry of these panels. We observe the largest improvement for small~$p^{(1)}$ and large~$p^{(2)}$ which is exactly when the combination shown in~\eqref{eq:multitracer-bias-scaling} is maximized. In this case, the double-tracer approach can improve the constraints by as much as a factor of two over a single-tracer analysis which is much larger than what we saw for fixed~$p^{(i)}$ and varying~$b_1^{(i)}$ in the left panels. Figure~\ref{fig:multi-p}%
displays the constraining power as a function of the scaling exponent~$\Delta$ for a few pairs of tracers characterized by different values of~$p^{(i)}$.\footnote{We also tested the possibility of using galaxy samples for which the parameter~$p^{(i)}$ is redshift dependent. The results shown here are robust to these changes.} This is further confirmation of our findings and we explicitly see that maximizing the combination of biases given in~\eqref{eq:multitracer-bias-scaling} leads to the same relative improvement in~$\sigma(\fnl^\Delta)$ for all~$\Delta$.\medskip

The results shown in the previous figures suggest that it may be possible to improve~$\sigma(\fnl^\Delta)$ by more than a factor of two if we had knowledge about the parameter~$p^{(i)}$ of the tracers. As previously discussed, however, most forecasts for current or near-future galaxy surveys assume the scale-dependent bias takes the form predicted by a universal halo mass function. In particular, these surveys do not select targets based on the value of~$p$~(or~$b_\phi$) which implies that there is some level of uncertainty in the actual value of~$p$ for the employed galaxy samples. Since the parameter~$p$ is degenerate with~$\fnl^\Delta$, this uncertainty in~$p$ could complicate the measurement of~$\fnl^\Delta$. While complete ignorance of~$p$~(i.e.\ marginalizing over it as a free parameter in each redshift bin) would make constraining~$\fnl^\Delta$ impossible, it would be helpful to know the required level of precision so that the universality assumption does not significantly affect the inference~of~$\fnl^\Delta$.

\begin{figure}
	\centering
	\includegraphics{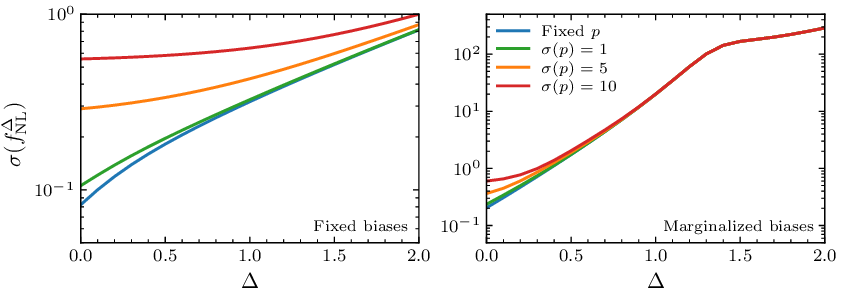}\vspace{-7pt}
	\caption{Forecast for $\sigma(\fnl^\Delta)$ for different Gaussian priors~$\sigma(p)$ on the parameter~$p$ in the universality relation~\eqref{eq:universality} for the billion-object survey~(at fixed $\Lambda$CDM~parameters and with fiducial value $\bar{f}_\mathrm{NL}^\Delta = 1$ so that~$p$ can be independently varied). While marginalizing over~$p$ with wide priors significantly impacts the constraints for small~$\Delta$ in the case of fixed bias parameters~(\textit{left}), the realistic case of marginalizing over the biases~(\textit{right}) shows a relatively small degradation even for very wide priors on~$p$.}
	\label{fig:p}
\end{figure}

We investigate this question by imposing different Gaussian priors on the parameter~$p$ for a fiducial value of $p = 1$. Figure~\ref{fig:p} shows that an uncertainty in~$p$ mostly affects $\Delta \lesssim 0.5$ when marginalization over the biases. In addition, the degradation in the constraining power on~$\fnl^\Delta$ is limited to within a factor of less than three~(two) for an extremely conservative prior with width $\sigma(p) = 10\text{ (5)}$. This implies that a lack of knowledge of the precise value of~$p$ for the given galaxy samples does not significantly affect the forecasts. On the other hand, a careful consideration of the values of~$p$ in target selection could potentially lead to important improvements in~$\sigma(\fnl^\Delta)$.

\subsection{Multi-Tracer LSST and Comparison to SPHEREx}
\label{sec:LSST_multi}

So far, we have seen that the multi-tracer forecasts offer somewhat limited potential for improvements in the context of both current and future, realistic spectroscopic surveys. Acquiring spectra takes more observational time than imaging alone which makes it harder for spectroscopic surveys to reach the high number densities needed for sample-variance cancellation to be most effective. In principle, photometric surveys, such as~LSST, are therefore not limited in the same way, as indicated in Fig.~\ref{fig:nbar}, and have the potential to benefit significantly from a multi-tracer analysis. Following the LSST~science book~\cite{LSSTScience:2009jmu}, our LSST~forecasts however use a single tracer as shown in Table~\ref{tab:lsst-specs} and, consequently, do not capture \mbox{the full potential of this survey which we will now investigate}.\medskip

\begin{figure}
	\centering
	\includegraphics{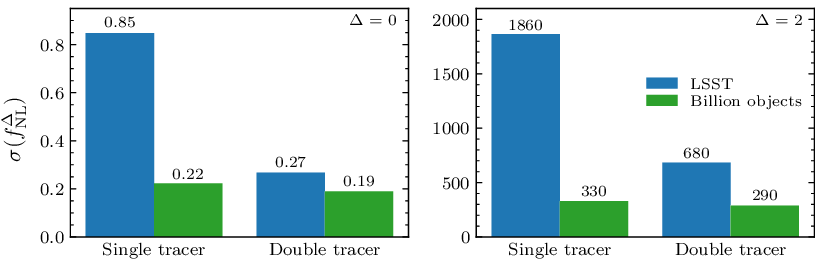}\vspace{-8pt}
	\caption{Comparison of the constraining power of single- and double-tracer versions of~LSST and the billion-object survey~(with fixed $\Lambda$CDM~parameters) for $\Delta = 0$~(\textit{left}) and~$2$~(\textit{right}). We can clearly see the significant improvement that a double-tracer analysis of~LSST could provide due to its large number density for the entire range of scaling exponents~$\Delta$.}\vspace{-6pt}
	\label{fig:lsst-billion-comparison}
\end{figure}

In order to explore the potential for an LSST~multi-tracer analysis, we consider a simplified case in which we split the LSST~sample of Table~\ref{tab:lsst-specs} into two subsamples of equal number density but different linear biases~$b_1^{(i)}$.\footnote{For simplicity, we split the LSST~sample in the same way as the sample for the billion-object survey, i.e.\ we take $b_1^{(1,2)}(z=0) = b_1^\mathrm{single}(z=0) \pm 0.4$ with equal number densities, $\bar{n}_g^{(1)} = \bar{n}_g^{(2)}$, to retain the single-tracer LSST~sample when combined. We refer to~\cite{MoradinezhadDizgah:2017szk} for similar forecasts for an LSST~sample split into ``blue'' and ``red'' galaxies.} In Figure~\ref{fig:lsst-billion-comparison},%
we compare the constraints~$\sigma(\fnl^\Delta)$ from single- and double-tracer configurations of~LSST and the billion-object survey. As for all of our spectroscopic surveys with their given number densities, there is a relatively small change in the constraining power for the billion-object survey.\footnote{In this context, it is important to remember the large maximum redshift for this survey~($\zmax = 5$) which results in a number density more or less comparable to DESI/Euclid for many redshift bins. The same number of objects in a smaller volume, as would for example be the case for a spectroscopic follow-up survey to~LSST with $\zmax = 3$, would not have the same limitation, but its effectiveness for most other science targets would be reduced~\cite{Chang:2022lrw}.} On the other hand, LSST~shows a remarkable improvement which is consistent with the factor of roughly three seen in Fig.~\ref{fig:nbar}. This increase in sensitivity however depends of course on the biases of the samples and would be diminished to less than a factor of two if we had instead taken $b_1^{(1,2)}(z=0) = b_1^\mathrm{single}(z=0) \pm 0.2$, for instance. We additionally note that the much larger number density of the LSST~samples appears to outweigh the larger number of (linear)~modes in the billion-object survey.

In this regard, LSST~has much more potential for testing inflationary physics than is typically seen in forecasts. Of course, spectroscopic information may prove essential in eliminating systematics or reliably splitting the LSST~sample. The improvements that we see in Fig.~\ref{fig:lsst-billion-comparison} however strongly encourage a future investigation that accounts for systematic effects. In addition, we have not included the information from cosmic-shear observations, which are sensitive to primordial non-Gaussianity in their own right~(see e.g.~\cite{Anbajagane:2023wif}) and may be useful in conjunction with the galaxy power spectra considered here. Cross-correlations with other observables, such as CMB~lensing and the kinetic Sunyaev-Zel'dovich effect, would additionally provide complementary constraining power~\cite{Schmittfull:2017ffw, Munchmeyer:2018eey, Chen:2021vba}.\medskip\vspace{-2pt}

Given the potential of~LSST as a multi-tracer survey, we should also revisit how it compares to other near-term surveys. We noted in~\textsection\ref{sec:information} that LSST and~SPHEREx result in similar forecasts despite having different strengths and weaknesses in their designs. SPHEREx~is a spectro-photometric survey with a larger sky coverage, while LSST~is a photometric survey with a larger number density. The comparison in~\textsection\ref{sec:information} is additionally complicated by the fact that we treated~LSST as a single-tracer survey following~\cite{LSSTScience:2009jmu} while multi-tracer forecasts would maximize the benefit of the higher number density.

We might have naively imagined that the slight advantage of~SPHEREx was due to the smaller photometric redshift error of most of its samples. (Specifically, the redshift uncertainty of~LSST is assumed to be $\sigma_{z0} = 0.05$, while the SPHEREx~samples are binned with maximum errors of $\sigma_{z0} = \{0.003, 0.01, 0.03, 0.1, 0.2\}$.) This is however not what is found in our forecasts. In Figure~\ref{fig:photoz}, we see that the potential impact of photometric redshift errors on~$\fnl^\Delta$ for $\Delta \lesssim 1.2$ is at most~$50\%$. At the same time, we observed in Fig.~\ref{fig:lsst-billion-comparison} that the potential improvements in~LSST from splitting the single sample into two is potentially a factor of three or more. This suggests that the number density is a larger effect than the quality of the photometric redshifts.\footnote{While the observed volume of~SPHEREx is in principle more than twice as large~(see Appendix~\ref{app:forecasting}), its number densities are much smaller, especially at high redshifts, which results in an effective volume that is smaller by a factor of almost four compared to~LSST.} At the same time, there are however many other factors that differ between these surveys and we would like to identify the most important individual design factors that set the two surveys apart~(at least for the purpose of constraining~$\fnl^\Delta$ from the scale-dependent bias in the power spectrum).

\begin{figure}
	\centering
	\includegraphics{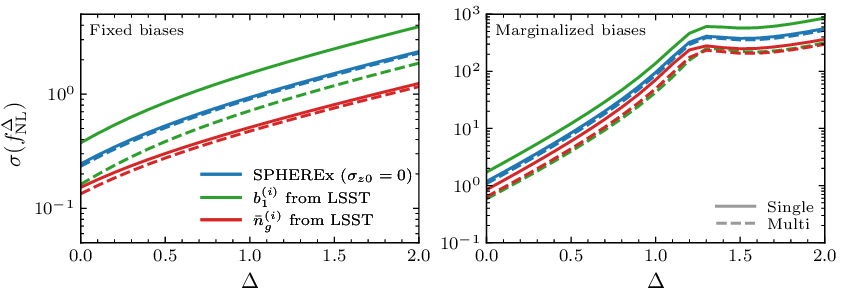}\vspace{-7pt}
	\caption{Constraining power of~SPHEREx with its fiducial biases~$b_1^{(i)}$ and number densities~$\bar{n}_g^{(i)}$ compared to a survey with either of these observational design factors replaced by LSST-like numbers for fixed~(\textit{left}) and marginalized~(\textit{right}) Gaussian bias parameters. The solid/dashed lines show the forecasts for a single-/multi-tracer version of these surveys~(five samples for~SPHEREx and two samples when considering LSST-like design factors) for which we ignore the photometric redshift errors in order to directly infer the sizable impact of these design choices.}
	\label{fig:spherex-lsst}
\end{figure}

In order to isolate the differences and control the individual factors in survey design, we compare~LSST and~SPHEREx by their biases and number densities in Fig.~\ref{fig:spherex-lsst},%
while ignoring the photometric redshift errors for both surveys, $\sigma_{z0} = 0$. To be precise, we take~SPHEREx as the fiducial survey and substitute either the linear biases or the number densities from a survey equivalent to~LSST~(see Appendix~\ref{app:forecasting} for details). When we keep the number densities of~SPHEREx and use the LSST-like biases, i.e.\ two samples with equal number densities, which add to the total number density of~SPHEREx, and the same biases as in our earlier LSST~forecasts just over the larger redshift range of~SPHEREx with the same scaling proportional to~$D(z)^{-1}$, SPHEREx~still outperforms in the single-tracer case, but underperforms in the double-tracer scenario. In this sense, SPHEREx~has a built-in advantage from the fiducial biases used in our forecasts, while our relatively large splitting in the linear biases of the LSST-like samples leads to a considerable improvement in the regime of sample-variance cancellation. If we instead keep the biases of~SPHEREx and use the number densities of~LSST, i.e.\ we include objects with LSST-like number densities according to~\eqref{eq:Ng_LSST} beyond its~$\zmax$, the latter becomes more sensitive. In fact, this advantage becomes larger when we additionally allow for a double-tracer configuration of~LSST, which takes full advantage of the difference in number densities. As a result, we should understand that the slightly better sensitivity of~SPHEREx indicated in Fig.~\ref{fig:future_surveys} is primarily due to the single-tracer nature of the LSST~forecast and the larger biases of the SPHEREx~samples.\medskip

Before concluding, we have to however state that the largest challenge for both~LSST and~SPHEREx will of course be identifying and mitigating the impact of large-angle~(and other potential) systematics. In this regard, the survey which will be most sensitive to~$\fnl^\Delta$ cannot be easily anticipated ahead of time. Either way, both surveys have the potential to in principle place constraints on general~PNG from the galaxy power spectrum alone~(i.e.\ without the galaxy bispectrum) that exceed those inferred from the Planck bispectrum for $\Delta \lesssim 1$.

\section{Conclusions and Outlook}
\label{sec:conclusions}

Primordial non-Gaussianity from scale-dependent bias offers one of the best avenues to test the physics of inflation in the coming decade and beyond. The signal imprinted by light particles coupled to the inflaton is localized on relatively large scales where analyses are less influenced by gravitational nonlinearities, baryonic physics and modeling errors. Equally importantly, the scaling behavior of this signal encodes the mass of these light particles, offering a unique window into the spectrum and interactions of particles during the inflationary epoch.\medskip

Pragmatically, the scale-dependent bias offers a method to test for new physics within the power spectra measured in large-scale structure. One might expect that the best measurements of primordial physics should ultimately come from higher-$N$-point statistics or even from analyses performed directly at the level of the observed maps.\footnote{In some cases, it has already been shown that the signals of extra fields are best understood as map-level features~\cite{Munchmeyer:2019wlh, Biagetti:2020skr, Baumann:2021ykm, Biagetti:2022qjl, Andrews:2022nvv, Jung:2022rtn} that may carry more signal to noise than low-order statistics.} Technical obstacles have however slowed down progress and obscured the ultimate sensitivity of LSS~surveys. In fact, analyses of the bispectrum only recently became possible for single-field inflation~(equilateral~PNG) and remain far from the demonstrated sensitivity of the~CMB. The established theoretical and observational control of LSS~power spectra therefore offer a reliable approach for near-term surveys to impact our understanding of inflation.\medskip

Previous work has largely focused on local~PNG for which the signal to noise is dominated by only the largest angular scales accessible in a given survey. These measurements are exciting since they have the potential to exceed the sensitivity of the~CMB and explore the interesting theoretical threshold of $\fnlloc \sim O(1)$. The much richer phenomenology of inflation however raises the question if optimizing a survey for~$\fnlloc$ is the same as optimizing for understanding inflation more broadly. For example, degeneracies~\cite{LoVerde:2014rxa, Vagnozzi:2018pwo, Shiveshwarkar:2023xjv} and large-angle~(systematic) effects~\cite{Pullen:2012rd, Alonso:2015uua, Castorina:2020blr, Castorina:2021xzs, Martinez-Carrillo:2021lcn, Foglieni:2023xca} that limit the local~PNG measurement may not impact other changes to the statistics. Similarly, some variations in the observational strategy or target selection may not significantly affect the constraining power on~$\fnlloc$, but could yield a vast increase in the insights that we can gain on other inflationary observables.\medskip

In order to extend the reach of current and future surveys to these compelling inflationary targets, we studied primordial non-Gaussianity associated with light fields during inflation over a wide range of their masses. We found that surveys optimized for~$\fnlloc$ tend to perform equally well over a broader mass range beyond the massless~(local) limit. Furthermore, we placed new constraints on these types of~PNG by analyzing the BOSS~DR12 dataset in two different and consistent ways. These results are consistent with previous analyses in the case of massless fields. In addition, they extend to heavier fields and other types of non-Gaussianity. They also show that bounds on these general non-Gaussian shapes can be simply inferred from scale-dependent bias measurements of~$\fnlloc$ and~$\fnleq$ if they are consistent with zero. Our measured constraints from BOSS~data are also consistent with our BOSS~forecasts which suggests that our understanding of the constraining power of these surveys, which we described in detail, is reflected in real data. These insights are likely relevant to the analysis strategies for current or near-term observations and the design of future surveys. For instance, we demonstrated the significant improvements in sensitivity that may be available from a multi-tracer analysis of LSST~data.\medskip

More broadly speaking, additional fields present during inflation are interesting in their own right, far beyond their potential impact on the galaxy power spectrum. While we focused on models leading to power-law corrections to the power spectrum, oscillatory contributions to inflationary spectra can arise from (more)~massive fields~\cite{Noumi:2012vr, Arkani-Hamed:2015bza, Lee:2016vti} or chemical potentials~\cite{Behbahani:2012be, Bodas:2020yho}, for instance. This vast range of phenomenology therefore indicates that there are many opportunities in the analysis of galaxy surveys~\cite{Meerburg:2016zdz, MoradinezhadDizgah:2017szk} that remain under-explored and, as we also show in this paper, minor additional work could result in much broader and deeper insights into the primordial universe.

\vskip20pt
\paragraph{Acknowledgments}
We are grateful to Alexandre Barreira, Daniel Baumann, Matteo Biagetti, Kyle Dawson, Olivier Dor\'e, Simone Ferraro, Marilena LoVerde, Azadeh Moradinezhad Dizgah, and Charuhas Shiveshwarkar for helpful discussions. The authors were supported by the US~Department of Energy under Grant~\mbox{DE-SC0009919}. B.\,W.~also acknowledges support from the Swedish Research Council~(Contract No.~\mbox{638-2013-8993}). Part of this work is based on observations obtained by the Sloan Digital Sky Survey~III (\mbox{SDSS-III}, \href{http://www.sdss3.org/}{http:/\!/www.sdss3.org/}). Funding for SDSS-III has been provided by the Alfred P.~Sloan Foundation, the Participating Institutions, the National Science Foundation and the US~Department of Energy Office of Science. We acknowledge the use of \texttt{CAMB}~\cite{Lewis:1999bs}, \texttt{CLASS}~\cite{Blas:2011rf}, \texttt{CLASS-PT}~\cite{Chudaykin:2020aoj}, \texttt{FAST-PT}~\cite{McEwen:2016fjn}, \texttt{IPython}~\cite{Perez:2007ipy} and \texttt{MontePython}~\cite{Audren:2012wb, Brinckmann:2018cvx}, and the Python packages \texttt{Matplotlib}~\cite{Hunter:2007mat}, \texttt{NumPy}~\cite{Harris:2020xlr} and~\texttt{SciPy}~\cite{Virtanen:2019joe}.

\clearpage
\appendix
\section{Forecasting Details}
\label{app:forecasting}

In this appendix, we collect additional information on our LSS~Fisher forecasts. We first provide the full set of definitions underlying our model for the galaxy power spectrum and then describe the experimental specifications of all galaxy surveys employed in the main text.

\subsection{Galaxy Power Spectrum Model}

We model the galaxy power spectrum at one-loop order in standard Eulerian perturbation theory and in the bias expansion where the galaxy overdensity~$\delta_g$ is a function of terms up to third order in the linear matter density contrast~$\delta_m$~[cf.~\eqref{eq:galaxy-overdensity}]. The theoretical galaxy power spectrum for two biased tracers~$A$ and~$B$ is then given by~\eqref{eq:Pg-th}. The various loop contributions in that equation are defined as follows:
\begin{align}
	\sigma^4				&= \int_{\q} P_\mathrm{lin}^2(q) \, ,	\\
	P_{22}(k)				&= 2 \int_{\q} \left[ F^{(s)}_2(\q,\k-\q) \right]^2 P_\mathrm{lin}(q) P_\mathrm{lin}(|\k-\q|) \, ,	\\
	P_{13}(k)				&= 6 P_\mathrm{lin}(k) \int_{\q} F^{(s)}_3(\k,\q,-\q) P_\mathrm{lin}(q) \, ,	\\
	P_{\delta^2}(k)			&= 2 \int_{\q} F^{(s)}_2(\q,\k-\q) P_\mathrm{lin}(q) P_\mathrm{lin}(|\k-\q|) \, ,	\\
	P_{s^2}(k)				&= 2 \int_{\q} F^{(s)}_2(\q,\k-\q) \left( \mu_-^2-\tfrac{1}{3} \right) P_\mathrm{lin}(q) P_\mathrm{lin}(|\k-\q|) \, ,	\\
	P_{\delta^2\delta^2}(k)	&= 2 \int_{\q} P_\mathrm{lin}(q) P_\mathrm{lin}(|\k-\q|) \, ,	\\
	P_{\delta^2s^2}(k)		&= 2 \int_{\q} \left( \mu_-^2 - \tfrac{1}{3} \right) P_\mathrm{lin}(q) P_\mathrm{lin}(|\k-\q|) \, ,	\\
	P_{s^2s^2}(k)			&= 2 \int_{\q} \left( \mu_-^2 - \tfrac{1}{3} \right)^{\!2} P_\mathrm{lin}(q) P_\mathrm{lin}(|\k-\q|) \, ,	\\
	P_{\Pi\Pi^{[2]}}(k)		&= 2P_\mathrm{lin}(k) \int_{\q} \left\{ \frac{2}{7} \left[\!\left(\frac{\k\cdot\q}{k\hskip1pt q}\right)^{\!2} - 1 \right] \frac{2}{3} \mu_-^2 + \frac{8}{63} \right\} P_\textrm{lin}(q) \, ,
\end{align}
with $\int_{\q} \equiv \int\hskip-3pt\frac{\d^3 q}{(2\pi)^3}$ and $\mu_- \equiv \frac{\q\cdot(\k-\q)}{q |\k-\q|}$. The functions~$F_2^{(s)}(\k_1,\k_2)$ and~$F_3^{(s)}(\k_1,\k_2,\k_3)$ are the symmetric, second- and third-order kernels of~$\delta_m$ in standard perturbation theory~\cite{Bernardeau:2001qr},
\begin{align}
	F_2^{(s)}(\k_1,\k_2)		=&\ \frac{5}{7} + \frac{2}{7}\left(\frac{\k_1\cdot\k_2}{k_1 k_2}\right)^{\!2} + \frac{1}{2}\frac{\k_1\cdot\k_2}{k_1 k_2}\left(\frac{k_1}{k_2} + \frac{k_2}{k_1}\right) \, ,	\\
	F_3^{(s)}(\k_1,\k_2,\k_3)	=&\ \frac{k^2}{27}\left[ \frac{\k_1\cdot(\k_2+\k_3)}{k_1^2(\k_2+\k_3)^2} G_2^{(s)}(\k_2,\k_3) + 2~\text{cyclic} \right]	\nonumber \\
								 &+ \frac{7}{54}\k\cdot \left[ \frac{\k_1+\k_2}{(\k_1+\k_2)^2} G_2^{(s)}(\k_1,\k_2) + 2~\text{cyclic} \right]	\\
								 &+ \frac{7}{54}\k\cdot \left[ \frac{\k_1}{k_1^2} F_2^{(s)}(\k_2,\k_3) + \text{2 cyclic} \right] ,	\nonumber
\end{align}
where $\k \equiv \k_1+\k_2+\k_3$ and $G_2^{(s)}(\k_1,\k_2)$ is the symmetric, second-order kernel of the velocity divergence~$\theta_m$,
\begin{equation}
	G_2^{(2)}(\k_1,\k_2) = \frac{3}{7} + \frac{4}{7}\left(\frac{\k_1\cdot\k_2}{k_1 k_2}\right)^{\!2} + \frac{1}{2}\frac{\k_1\cdot\k_2}{k_1 k_2}\left(\frac{k_1}{k_2} + \frac{k_2}{k_1}\right) .
\end{equation}

\subsection{Survey Specifications}

In the following, we provide detailed information about the spectroscopic and photometric redshift surveys that we employ in our Fisher forecasts, including the assumed redshift distribution in bias and number density.

\subsubsection*{Spectroscopic Surveys}

The experimental specifications of the spectroscopic redshift surveys~BOSS, DESI, Euclid, SPHEREx, MegaMapper and the billion-object survey are provided in Tables~\ref{tab:boss-specs} to~\ref{tab:billion-specs}.%
\begin{table}[b]
	\centering
	\begin{tabular}{l S[table-format=1.3] S[table-format=1.3]}
			\toprule
		$\bar{z}$													& 0.35	& 0.625	\\
			\midrule[0.065em]
		$b_1$ 														& 1.634	& 1.877	\\
		$\num{e3}\bar{n}_g\ [\si{\h\tothe{3}\per\Mpc\tothe{3}}]$	& 0.275	& 0.142	\\
		$V\ [\si{\per\h\tothe{3}\Gpc\tothe{3}}]$					& 2.18	& 4.15	\\
			\bottomrule
	\end{tabular}
	\caption{Basic specifications for a BOSS-like survey~\cite{Beutler:2019ojk}~(inspired by~\cite{BOSS:2016hvq} as detailed in~\cite{Baumann:2017gkg}), covering a sky area of~\SI{10252}{deg^2} with a total of about \num{1.2e6}~objects in a volume of roughly~\SI{6.3}{\per\h\tothe{3}\Gpc\tothe{3}}.}
	\label{tab:boss-specs}
	\bigskip
	\centering
	\begin{tabular}{l S[table-format=2.4] S[table-format=2.3] S[table-format=1.3] S[table-format=1.3] S[table-format=1.4] S[table-format=1.3] S[table-format=1.4] S[table-format=1.4] S[table-format=1.4]}
			\toprule
		$\bar{z}$ 															& 0.05	& 0.15	& 0.25	& 0.35	& 0.45	& 0.65	& 0.75	& 0.85	& 0.95	\\
			\midrule[0.065em]
		$b_1$																& 1.40	& 1.48	& 1.55	& 1.61	& 1.67	& 2.05	& 1.71	& 1.71	& 1.53	\\
		$\num{e3}\bar{n}_g\ [\si{\h\tothe{3}\per\Mpc\tothe{3}}]$\hskip-9pt~	& 38.8	& 15.7	& 3.96	& 0.883	& 0.0992& 0.591	& 1.31	& 0.920	& 0.779	\\
		$V\ [\si{\per\h\tothe{3}\Gpc\tothe{3}}]$							& 0.0356& 0.229	& 0.560	& 0.979	& 1.44	& 2.39	& 2.83	& 3.24	& 3.61	\\
			\bottomrule
	\end{tabular}
	\begin{tabular}{l S[table-format=2.4] S[table-format=2.3] S[table-format=1.3] S[table-format=1.3] S[table-format=1.4] S[table-format=1.3] S[table-format=1.4] S[table-format=1.4] S[table-format=1.4]}
			\toprule
		$\bar{z}$															& 1.05	& 1.15	& 1.25	& 1.35	& 1.45	& 1.55	& 1.65	& 1.75	& 1.85	\\
			\midrule[0.065em]
		$b_1$																& 1.45	& 1.48	& 1.47	& 1.47	& 1.69	& 1.68	& 2.27	& 2.45	& 2.47	\\
		$\num{e3}\bar{n}_g\ [\si{\h\tothe{3}\per\Mpc\tothe{3}}]$\hskip-9pt~	& 0.466	& 0.398	& 0.387	& 0.180	& 0.133	& 0.110	& 0.0387& 0.0197& 0.0208\\
		$V\ [\si{\per\h\tothe{3}\Gpc\tothe{3}}]$							& 3.94	& 4.24	& 4.49	& 4.71	& 4.90	& 5.05	& 5.18	& 5.29	& 5.37	\\
			\bottomrule
	\end{tabular}
	\caption{Basic specifications for~DESI~\cite{Beutler:2019ojk}~(derived from~\cite{DESI:2016fyo} as explained in~\cite{Baumann:2017gkg}), covering a sky area of~\SI{14000}{deg^2} with a total of about \num{2.7e7}~objects in a volume of roughly~\SI{58}{\per\h\tothe{3}\Gpc\tothe{3}}.}
	\label{tab:desi-specs}
\end{table}
We provide the linear bias~$b_1$ and number density~$\bar{n}_g$ for each sample and redshift bin with mean redshift~$\bar{z}$ and spherical volume~$V$. For~DESI, different types of tracers were combined into one galaxy sample with a single effective number density and bias following~\cite{Baumann:2017gkg, Beutler:2019ojk}. Since the number density of~DESI is not in the regime where sample-variance cancellation is most effective (see Fig.~\ref{fig:nbar}), our results should be relatively insensitive to this choice. For all surveys, we treat each redshift bin as being independent so that the Fisher matrix for the entire survey is the sum of the Fisher matrices associated with each separate redshift bin.
\begin{table}
	\centering
	\begin{tabular}{l S[table-format=1.3] S[table-format=1.3] S[table-format=1.2] S[table-format=1.2] S[table-format=1.2] S[table-format=1.2] S[table-format=1.2] S[table-format=1.3]}
			\toprule
		$\bar{z}$													& 0.65	& 0.75	& 0.85	& 0.95	& 1.05	& 1.15	& 1.25	& 1.35	\\
			\midrule[0.065em]
		$b_1$														& 1.06	& 1.11	& 1.16	& 1.21	& 1.27	& 1.33	& 1.38	& 1.44	\\
		$\num{e3}\bar{n}_g\ [\si{\h\tothe{3}\per\Mpc\tothe{3}}]$	& 0.637	& 1.46	& 1.63	& 1.50	& 1.33	& 1.14	& 1.00	& 0.837	\\
		$V\ [\si{\per\h\tothe{3}\Gpc\tothe{3}}]$					& 2.56	& 3.03	& 3.47	& 3.87	& 4.23	& 4.54	& 4.81	& 5.05	\\
			\bottomrule
	\end{tabular}
	\begin{tabular}{l S[table-format=1.3] S[table-format=1.3] S[table-format=1.3] S[table-format=1.3] S[table-format=1.3] S[table-format=1.4] S[table-format=1.4]}
			\toprule
		$\bar{z}$													& 1.45	& 1.55	& 1.65	& 1.75	& 1.85	& 1.95	& 2.05	\\
			\midrule[0.065em]
		$b_1$														& 1.51	& 1.54	& 1.63	& 1.70	& 1.85	& 1.90	& 1.26	\\
		$\num{e3}\bar{n}_g\ [\si{\h\tothe{3}\per\Mpc\tothe{3}}]$	& 0.652	& 0.512	& 0.357	& 0.246	& 0.149	& 0.0904& 0.0721\\
		$V\ [\si{\per\h\tothe{3}\Gpc\tothe{3}}]$					& 5.25	& 5.41	& 5.55	& 5.67	& 5.76	& 5.83	& 5.88	\\
			\bottomrule
	\end{tabular}\vspace{-2pt}
	\caption{Basic specifications for the Euclid survey~\cite{Baumann:2017gkg}~(derived from~\cite{Font-Ribera:2013rwa}), covering a sky area of~\SI{15000}{deg^2} with a total of about \num{4.9e7}~objects in a volume of roughly~\SI{71}{\per\h\tothe{3}\Gpc\tothe{3}}.\vspace{-4pt}}
	\label{tab:euclid-specs}
\end{table}
\begin{table}
	\centering
	\begin{tabular}{l c c c c c c c c c c c}
			\toprule
		$\bar{z}$					& 0.1	& 0.3	& 0.5	& 0.7	& 0.9	& 1.3	& 1.9	& 2.5	& 3.1	& 3.7	& 4.3	\\
			\midrule[0.065em]
		$V$							& 0.584	& 3.40	& 7.43	& 11.5	& 15.1	& 60.4	& 71.4	& 73.6	& 71.8	& 68.3	& 64.2	\\
			\midrule[0.065em]
		$b_1^{(1)}$					& 1.3	& 1.5	& 1.8	& 2.3	& 2.1	& 2.7	& 3.6	& 2.3	& 3.2	& 2.7	& 3.8	\\
		$\num{e5}\bar{n}_g^{(1)}$	& 997	& 411	& 50.1	& 7.05	& 3.16	& 1.64	& 0.359	& 0.0807& 0.184	& 0.150	& 0.113	\\
			\midrule[0.065em]
		$b_1^{(2)}$					& 1.2	& 1.4	& 1.6	& 1.9	& 2.3	& 2.6	& 3.4	& 4.2	& 4.3	& 3.7	& 4.6	\\
		$\num{e5}\bar{n}_g^{(2)}$	& 1230	& 856	& 282	& 93.7	& 43.0	& 5.00	& 0.803	& 0.383	& 0.328	& 0.107	& 0.0679\\
			\midrule[0.065em]
		$b_1^{(3)}$					& 1.0	& 1.3	& 1.5	& 1.7	& 1.9	& 2.6	& 3.0	& 3.2	& 3.5	& 4.1	& 5.0	\\
		$\num{e5}\bar{n}_g^{(3)}$	& 1340	& 857	& 362	& 294	& 204	& 21.2	& 0.697	& 0.202	& 0.143	& 0.193	& 0.0679\\
			\midrule[0.065em]
		$b_1^{(4)}$					& 0.98	& 1.3	& 1.4	& 1.5	& 1.7	& 2.2	& 3.6	& 3.7	& 2.7	& 2.9	& 5.0	\\
		$\num{e5}\bar{n}_g^{(4)}$	& 2290	& 1290	& 535	& 495	& 415	& 79.6	& 7.75	& 0.787	& 0.246	& 0.193	& 0.136	\\
			\midrule[0.065em]
		$b_1^{(5)}$					& 0.83	& 1.2	& 1.3	& 1.4	& 1.6	& 2.1	& 3.2	& 4.2	& 4.1	& 4.5	& 5.0	\\
		$\num{e5}\bar{n}_g^{(5)}$	& 1490	& 752	& 327	& 250	& 183	& 73.4	& 25.3	& 5.41	& 2.99	& 0.941	& 0.204	\\
			\bottomrule
	\end{tabular}\vspace{-2pt}
	\caption{Basic specifications for SPHEREx~\cite{SPHEREx:2020git}, covering a sky fraction $\fsky = 0.75$ with a total of about \num{7.0e8}~objects in a volume of roughly~\SI{450}{\per\h\tothe{3}\Gpc\tothe{3}}. SPHEREx~is a spectro-photometric survey and the observed objects are divided into five samples based on their photometric redshift uncertainty bin, with respective maximum error $\sigma_{z0} = \{0.003, 0.01, 0.03, 0.1, 0.2\}$. The volume~$V$ and the galaxy number density~$\bar{n}_g$ are given in units of~\si{\per\h\tothe{3}\Gpc\tothe{3}} and~\si{\h\tothe{3}\per\Mpc\tothe{3}}, respectively.\vspace{-3pt}}
	\label{tab:spherex-specs}
\end{table}
\begin{table}
	\centering
	\begin{tabular}{l S[table-format=2.3] S[table-format=2.3] S[table-format=2.3] S[table-format=2.3]}
			\toprule
		$\bar{z}$													& 2.0	& 3.0	& 4.0	& 5.0	\\
			\midrule[0.065em]
		$b_1$														& 2.5	& 4.0	& 3.5	& 5.5	\\
		$\num{e3}\bar{n}_g\ [\si{\h\tothe{3}\per\Mpc\tothe{3}}]$	& 0.98	& 0.12	& 0.10	& 0.040	\\
		$V\ [\si{\per\h\tothe{3}\Gpc\tothe{3}}]$					& 54.0	& 54.4	& 50.0	& 44.8	\\
			\bottomrule
	\end{tabular}
	\caption{Basic specifications for MegaMapper~\cite{Ferraro:2019uce}, covering a sky area of~\SI{14000}{deg^2} with a total of about \num{6.6e7}~objects in a volume of roughly~\SI{200}{\per\h\tothe{3}\Gpc\tothe{3}}.}
	\label{tab:megamapper-specs}
\end{table}
\begin{table}
	\centering
	\begin{tabular}{l S[table-format=1.2] S[table-format=2.2] S[table-format=2.2] S[table-format=2.2] S[table-format=2.2] S[table-format=2.2] S[table-format=2.2] S[table-format=2.2] S[table-format=2.2] S[table-format=2.2]}
			\toprule
		$\bar{z}$																	& 0.25	& 0.75	& 1.25	& 1.75	& 2.25	& 2.75	& 3.25	& 3.75	& 4.25	& 4.75	\\
			\midrule[0.065em]
		$V\ [\si{\per\h\tothe{3}\Gpc\tothe{3}}]$									& 4.78	& 20.6	& 32.8	& 38.8	& 40.8	& 40.7	& 39.4	& 37.7	& 35.9	& 34.0	\\
			\midrule[0.065em]
		$b_1^{(1)}$																	& 2.28	& 2.94	& 3.66	& 4.41	& 5.17	& 5.94	& 6.72	& 7.51	& 8.29	& 9.07	\\
		$\num{e3}\bar{n}_g^{(1)}\, [\si{\h\tothe{3}\per\Mpc\tothe{3}}]$\hskip-10pt~	& 1.54	& 1.54	& 1.54	& 1.54	& 1.54	& 1.54	& 1.54	& 1.54	& 1.54	& 1.54	\\
			\midrule[0.065em]
		$b_1^{(2)}$																	& 1.37	& 1.76	& 2.19	& 2.64	& 3.10	& 3.57	& 4.03	& 4.50	& 4.97	& 5.45	\\
		$\num{e3}\bar{n}_g^{(2)}\, [\si{\h\tothe{3}\per\Mpc\tothe{3}}]$\hskip-10pt~	& 1.54	& 1.54	& 1.54	& 1.54	& 1.54	& 1.54	& 1.54	& 1.54	& 1.54	& 1.54	\\
			\bottomrule
	\end{tabular}
	\caption{Basic specifications for our~(futuristic) spectroscopic billion-object survey, covering a sky fraction $\fsky = 0.5$ with a total of \num{e9}~objects in a volume of roughly~\SI{330}{\per\h\tothe{3}\Gpc\tothe{3}}. The observed objects are divided into two samples based on their linear bias, $b_1(z=0) = 2.0\text{ and }1.2$.}
	\label{tab:billion-specs}
\end{table}

\subsubsection*{Photometric Surveys}

To forecast Vera Rubin Observatory's~LSST, which is a photometric redshift survey, we use the ``gold'' sample defined in their science book~\cite{LSSTScience:2009jmu}, which includes more than four billion galaxies over~\SI{20000}{deg^2}. The number of objects in each redshift bin is calculated according to
\begin{equation}
	N_g(\zmin,\zmax) = \left.\frac{N_\mathrm{tot} \Omega_\mathrm{sky}}{2 z_0^2} \left(2z_0^2+2z_0 z + z^2\right) \ee^{-z/z_0} \right|^{z = \zmax}_{z = \zmin} ,	\label{eq:Ng_LSST}
\end{equation}
where $N_\mathrm{tot} = \SI{0.55}{arcmin^{-2}}$, $\Omega_\mathrm{sky} = \SI{20000}{deg^2}$ and $z_0 = 0.3$. The bias is taken to be $b_1(z) = 0.95/D(z)$ and we assume a photometric redshift error of $\sigma_{z0} = 0.05$. Table~\ref{tab:lsst-specs}%
\begin{table}[b]
	\centering
	\begin{tabular}{l S[table-format=3.3] S[table-format=3.3] S[table-format=3.3] S[table-format=3.3] S[table-format=3.3] S[table-format=3.3]}
			\toprule
		$\bar{z}$													& 0.125	& 0.375	& 0.625	& 0.875	& 1.125	& 1.375	\\
			\midrule[0.065em]
		$b_1$														& 1.01	& 1.16	& 1.31	& 1.48	& 1.65	& 1.82	\\
		$\num{e3}\bar{n}_g\ [\si{\h\tothe{3}\per\Mpc\tothe{3}}]$	& 292	& 183	& 108	& 63.6	& 36.8	& 21.0	\\
		$V\ [\si{\per\h\tothe{3}\Gpc\tothe{3}}]$					& 0.709	& 3.93	& 8.10	& 11.9	& 14.9	& 17.0	\\
			\bottomrule
	\end{tabular}
	\begin{tabular}{l S[table-format=3.3] S[table-format=3.3] S[table-format=3.3] S[table-format=3.3] S[table-format=3.3] S[table-format=3.3]}
			\toprule
		$\bar{z}$													& 1.625	& 1.875	& 2.125	& 2.375	& 2.625	& 2.875	\\
			\midrule[0.065em]
		$b_1$														& 2.00	& 2.18	& 2.36	& 2.54	& 2.73	& 2.91	\\
		$\num{e3}\bar{n}_g\ [\si{\h\tothe{3}\per\Mpc\tothe{3}}]$	& 11.9	& 6.46	& 3.55	& 1.94	& 0.969	& 0.646	\\
		$V\ [\si{\per\h\tothe{3}\Gpc\tothe{3}}]$					& 18.4	& 19.2	& 19.7	& 19.9	& 19.8	& 19.6	\\
			\bottomrule
	\end{tabular}
	\caption{Basic specifications for Vera Rubin Observatory's~LSST~\cite{LSSTScience:2009jmu}, covering a sky area of~\SI{20000}{deg^2} with a total of about \num{4e9}~objects in a volume of roughly~\SI{170}{\per\h\tothe{3}\Gpc\tothe{3}}. The photometric redshift error is $\sigma_{z0} = 0.05$.}
	\label{tab:lsst-specs}
\end{table}
explicitly provides the derived experimental specifications for this survey with $\zmax = 3$ and twelve redshift bins with width~$\Delta z = 0.25$.

\clearpage
\section{Measuring the Scaling Behavior}
\label{app:scaling}

We have assumed for all forecasts throughout the paper that the fiducial value of the non-Gaussian amplitude~$\fnl^\Delta$ vanishes. This is of course motivated by the fact that we have only been placing observational bounds on~PNG that are consistent with $\fnl^\Delta = 0$, including in Section~\ref{sec:analysis}. We additionally always considered the scaling exponent~$\Delta$ to be fixed. In this appendix, we will consider nonzero fiducial PNG~amplitudes and clarify to what degree measurements of the galaxy power spectrum are sensitive to the precise value of~$\Delta$.\medskip

\begin{figure}
	\centering
	\includegraphics{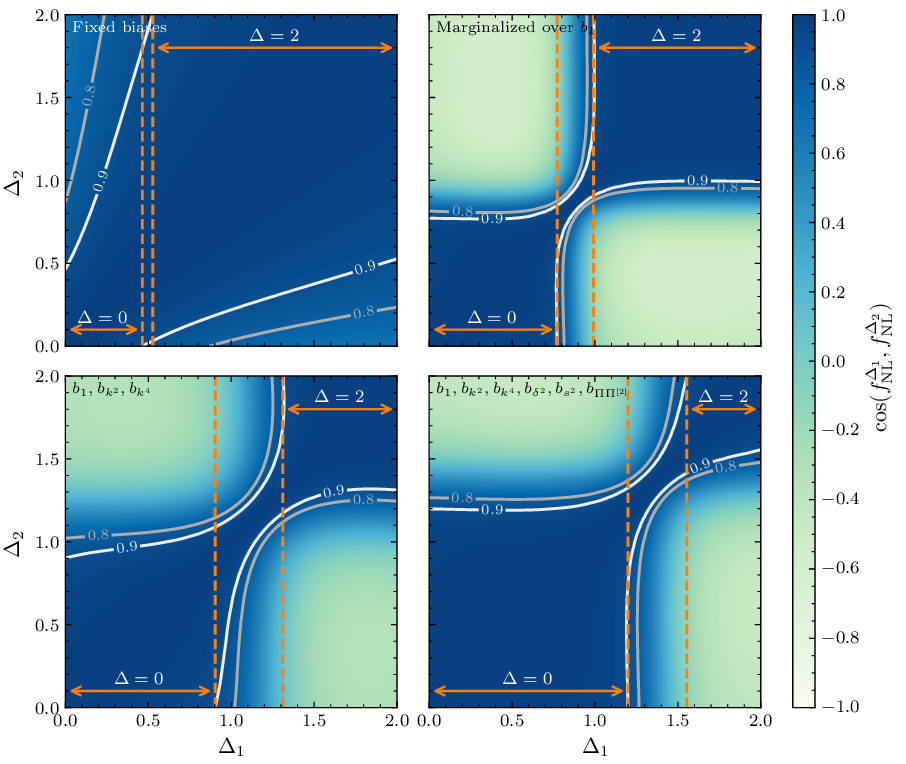}\vspace{-7pt}
	\caption{Correlation matrix for measurements of galaxy power spectra with different values of the non-Gaussian scaling exponent~$\Delta$ as defined in~\eqref{eq:cos-f1-f2} for the billion-object survey and four biasing models. The lower right panel is equivalent to Fig.~\ref{fig:cos-f1-f2}, which shows the correlations for~BOSS. As in that figure, we hold the $\Lambda$CDM~parameters fixed, take the fiducial non-Gaussian amplitudes to be zero, $\bar{f}_\mathrm{NL}^{\Delta_i} = 0$, and indicate the coverage of the local and equilateral templates with a correlation coefficient of at least~$0.9$ by the orange lines. We can clearly observe the impact of the biasing model and the different experimental specifications compared to~BOSS.}
	\label{fig:cos-f1-f2-full}
\end{figure}

We defined the correlation between the non-Gaussian signals of two different values of~$\Delta$ in~\eqref{eq:cos-f1-f2} in terms of the respective Fisher matrix elements~(after marginalization) for a fiducial value of $\bar{f}_\mathrm{NL}^{\Delta_{1,2}} = 0$. For BOSS, we can observe in Fig~\ref{fig:cos-f1-f2} that this correlation is close to unity, even for widely separated values of~$\Delta$. We see in Fig.~\ref{fig:cos-f1-f2-full}%
that this statement still holds for the billion-object survey. In fact, the range of~$\Delta$ which are highly correlated with the local shape further expanded. Naively, this suggests there is little information about the shape of the signal, described by~$\Delta$, encoded in these observables.

\begin{figure}
	\centering
	\includegraphics{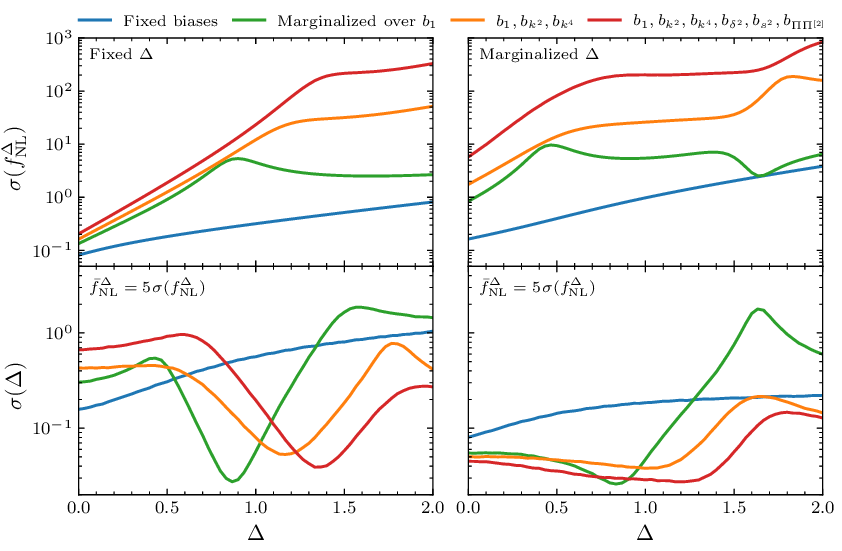}\vspace{-7pt}
	\caption{Forecasts for~$\sigma(\fnl^\Delta)$~(\textit{top}) and~$\sigma(\Delta)$~(\textit{bottom}) as a function of~$\Delta$ if the scaling exponent is either fixed~(\textit{left}) or varied~(\textit{right}) for the billion-object survey. To compute the constraint on~$\Delta$, we take the fiducial value of the non-Gaussian amplitude, $\bar{f}_\mathrm{NL}^\Delta$, to be the value necessary to achieve a $5\sigma$~detection in the respective forecast. We fix the $\Lambda$CDM~parameters and marginalize over four different biasing models.}
	\label{fig:5-sigma}
\end{figure}

To first approximation, the fiducial value of $\bar{f}_\mathrm{NL}^\Delta = 0$ is the reason why we see little ability to distinguish different values of~$\Delta$. Since there is no signal to measure, it is not surprising that the precise shape is not very important. If we were instead to detect~$\fnl^\Delta$, we would expect that the signal to noise would be a fairly sensitive function of~$\Delta$ and we could therefore measure the true value of~$\Delta$ precisely. This is exactly what is found in Fig.~\ref{fig:5-sigma},%
in which we show the forecasted constraints for measuring~$\Delta$,~$\sigma(\Delta)$, assuming that~$\fnl^\Delta$ is detected at a significance of~$5\sigma$ in the billion-object survey. The key result is that if we detect~$\fnl^\Delta$ while holding~$\Delta$ fixed, we expect to have enough sensitivity to measure the scaling exponent with $\sigma(\Delta) \sim 0.05 - 0.5$. In this sense, the strategy of searching for~$\fnl^\Delta$ at fixed~$\Delta$ is a reliable strategy to search for new physics, but would yield additional information about the origin of the signal even at the threshold of a detection. When we instead marginalize over~$\Delta$, we require a significantly larger non-Gaussian amplitude~$\fnl^\Delta$ to be detected. In this case, the corresponding uncertainties on~$\Delta$ would however be generally smaller, with $\sigma(\Delta) < 0.1$.

At first sight, it might seem surprising that the curves of~$\sigma(\Delta)$ in Fig.~\ref{fig:5-sigma} have minima at the values of~$\Delta$ where the curves of~$\sigma(\fnl^\Delta)$ have maxima or downward bends. By comparing the curves for different biasing models, we can however see that these features are the result of marginalizing over the bias parameters. For example, we noted in the main text that the maximum in~$\sigma(\fnl^\Delta)$ near $\Delta = 1$ when marginalizing over~$b_1$ arises because the linear bias~$b_1$ and~$\fnl^{\Delta=1}$ are degenerate at high~$k$. The fact that the measurement of~$\fnl^\Delta$ is less constraining at this point then also implies that we are very sensitive to the exact value of the scaling exponent~$\Delta$. When we hold~$\fnl^\Delta / \sigma(\fnl^\Delta)$ fixed, we should therefore expect that~$\sigma(\Delta)$ is smaller in regions of~$\Delta$ where~$\fnl^\Delta$ is degenerate with bias parameters.\medskip

Let us finally return to the correlation coefficients between scale-dependent biases with different scaling exponent~$\Delta$. In order to be consistent, the impact of marginalization over the bias parameters must also appear in~$\cos(\fnl^{\Delta_1}, \fnl^{\Delta_2})$. Specifically, the impact of the marginalization has a large impact on~$\sigma(\fnl^\Delta)$ for larger values of~$\Delta$. For the inferred constraints to be consistent with our forecasts, this has to imply that the correlation between large and small~$\Delta$ decreases by a similar factor. If this was not the case, our constraints inferred from~$\fnlloc$ at large~$\Delta$ would be stronger than directly measuring~$\fnl^\Delta$.

This expectation is precisely what occurs, as shown in Fig.~\ref{fig:cos-f1-f2-full}. The correlation coefficients for larger~$\Delta$ are affected by the number of bias parameters that are marginalized over. With more marginalized bias parameters, the range over which $\Delta < 2$ and $\Delta = 2$ are degenerate decreases significantly. This is expected since the (equilateral)~signal for $\Delta = 2$ gets most of its information from large wavenumbers, where the biases play a significant role~(cf.~\textsection\ref{sec:information}). In contrast, the increase of the range of strong overlap with $\Delta = 0$ is likely due to the increasing difficulty to distinguish similar shapes of~$\Delta$ with the inclusion of more bias parameters. Finally, it is also worth noticing that the non-Gaussian signal with $\Delta = 0$~(local~PNG) appears to be anti-correlated with the scale-dependent bias signal for $\Delta \sim 2$ in the last three panels of Fig.~\ref{fig:cos-f1-f2-full} in which we marginalize over the biasing model.

\clearpage
\phantomsection
\addcontentsline{toc}{section}{References}
\bibliographystyle{utphys}
\bibliography{references}

\end{document}